\documentclass[11pt,fleqn]{article}
\usepackage{latexsym,amsmath,amssymb,textgreek}
\usepackage{longtable,booktabs,adjustbox,multirow,multicol}
\usepackage{caption}
\usepackage{a4wide}
\usepackage{graphicx}
\usepackage{epstopdf}
\usepackage{xcolor}
\begin{document}
\numberwithin{equation}{section}
\renewcommand{\thefootnote}{*}
\renewcommand{\figurename}{\bf Fig.}
\thispagestyle{empty}

\begin{center}
{\Large \Large\bf On the order statistics from the XLindley distribution and associated inference with an application to fatigue data}\\
\vspace{.5cm}
{\bf Zuber Akhter$^1$\footnote{Corresponding author e-mail address: akhterzuber022@gmail.com},~S.M.T.K. MirMostafaee$^2$,~Abu Bakar$^3$,~ Ehsan Ormoz$^4$ }\\

\vspace{.3cm}
$^1$Department of Statistics\\
University of Delhi, Delhi-110 007, India

\vspace{.3cm}
$^2$Department of Statistics\\
University of Mazandaran, P.O. Box 47416-1467, Babolsar, Iran

\vspace{.3cm}
$^3$Department of Statistics and
Operations Research\\
Aligarh Muslim University, Aligarh-202 002, India

\vspace{.3cm}
$^4$Department of Statistics\\ Mashhad Branch, Islamic Azad University, Mashhad, Iran

\end{center}

\begin{abstract}
\noindent
In this paper, we consider the order statistics from a newly-introduced lifetime distribution, called the XLindley distribution. 
We have derived explicit closed form expressions for the single moments and product moments of order statistics from the XLindley distribution. Utilizing these expressions, we calculated the means, variances, and covariances of order statistics for sample sizes ranging from $n = 1$ to $n = 10$ and arbitrarily selected parameter values.  Additionally, these moments allow us to identify the best linear unbiased estimators and best linear invariant estimators   for the location and scale parameters, based on both complete samples and Type-II right censored samples. We also address the linear prediction of unobserved order statistics based on Type-II right-censored samples. 
We also explore the formulation of confidence intervals for both location and scale parameters, along with prediction intervals for unobserved order statistics.
To provide comparison and illustration, we conduct a simulation study and analyze a real data example.  Finally, we conclude with several remarks.

\end{abstract}
{\bf Keywords:} Order statistics; Best linear unbiased estimators; XLindley distribution; Best linear invariant estimators; Prediction.

\section{Introduction}
Order statistics play a crucial role in statistical theory, the probability distribution of the ordered values from a sample of random variables. The concept of order statistics extends to the arrangements of the sample observations in ascending or descending order. Order statistics have applications in reliability theory, engineering, quality control, environmental and medical sciences, economics, finance and auction theory. Order statistics have attracted the interest of many researchers among them David and Nagaraja \cite{david}, Balakrishnan and Rao \cite{BR} and Arnold et. al \cite{arnd}.

\vspace{0.25cm}
Lifetime data analysis is a crucial field within statistics, focusing on modeling and studying the time until failure or survival in various phenomena. The real-world applications of recent computational techniques in various fields like medicine, finance, biological engineering, and statistics. Statistical analysis depends significantly on the fundamental probability model or distributions. With its memoryless property and constant failure rate function, the exponential distribution is one of the most popular distributions for lifetime data. Sometimes, the exponential distribution might not suffice for representing certain data types showcasing various forms of failure rate functions, including increasing, decreasing, unimodal, or bathtub-shaped distributions. Therefore, several alternative distributions are proposed for the limitation of the exponential distribution which are more flexible and have a better fit for lifetime data.  The Lindley distribution \cite{lind} can be derived by the mixture of exponential($\psi$) and Gamma$(2,\psi)$. 
A generalization of the Lindley distribution with an extra shape parameter has been developed by Shanker and Mishra \cite{snker1}, who claimed that this distribution surpasses both the exponential and Lindley distributions, concerning failure rate and mean residual life functions. Shanker et al. \cite{snker2}, added a  parameter in the Lindley distribution to introduce the two-parameter Lindley distribution which is  more capable of fitting data than the Lindley distribution. 
The XLindley distribution is also a mixture of Exponential ($\psi$) and Lindley($\psi$) distributions, originally proposed by Chouia and Zeghdoudi \cite{Chouia},   being straightforward and easily applicable, offers simple formulas for computing key statistical measures such as mean, variance, coefficient of variation, skewness, kurtosis, and index of dispersion. It is a simpler distribution than  more complicated ones, and has been found  to provide a better fit for many real data  sets, for example
Corona, Ebola and Nipah virus data.

\vspace{.35cm}
 The probability density function (pdf) of the XLindley distribution with one parameter  $\psi$ is given by 
\begin{equation}\label{pdfxl}
\mathfrak{f}(x)=\frac{\psi^2 (2+\psi +x)}{(1+\psi)^2}{\rm e}^{-\psi x};\quad x>0;~\psi>0,
\end{equation}
\vspace{.2cm}
and therefore its cumulative distribution function (cdf) is
\begin{equation}\label{cdfxl}
\digamma(x)=1-\left( 1+\frac{\psi x }{ (1+\psi)^2}\right) {\rm e}^{-\psi{x}};\quad x>0;~\psi>0.
\end{equation}

Now, consider the three parameter  XLindley distribution (location and scale parameters added), so the pdf and cdf of the three parameter XLindley distribution are
\vspace{.2cm}
\begin{equation}\label{pdfsxl}
\mathfrak{f}(x;\varphi,\sigma,\psi)=\frac{ \psi^2 (2+\psi + (\frac{x-\varphi}{\sigma}))}{\sigma(1+\psi)^2}\exp\left(- \dfrac{\psi (x-\varphi)}{\sigma}\right);\quad x>\varphi,
\end{equation}
\vspace{.2cm}
and
\begin{equation} \label{cdfsxl}
\displaystyle \digamma(x;\varphi,\sigma,\psi)=1-\left( 1+\dfrac{\psi (x-\varphi) }{\sigma (1+\psi)^2}\right) \exp\left(- \dfrac{\psi (x-\varphi)}{\sigma}\right);\quad x>\varphi,
\end{equation}
respectively, 
where $\psi>0$, $\sigma>0$ and $\varphi \in \mathfrak{R}$ the shape, scale and location parameters, respectively. We write $X\sim XL(\varphi,\sigma,\psi)$ if the pdf of $X$ can written as (\ref{pdfsxl}).

\vspace{.2cm}
Chouia and Zeghdoudi \cite{Chouia} studied  various statistical properties and aspects of the XLindley distribution such as moments, skewness, kurtosis, quantile function, hazard rate function, maximum likelihood estimation, method of moments, and so on. This distribution has been the subject of numerous studies by other authors, who have explored its properties, characterization, statistical inference, and applications. Some notable examples include Alotaibi et. al \cite{Alo}, Metiri et. al \cite{Metri}, Nassar et. al \cite{Nsr}, and Zanjiran and MirMostafaee \cite{Zanjiran}.
The various applications of moments of order statistics are extensively documented in the statistical literature. These moments play a crucial role in statistical modeling, inference, decision-making methods, reliability analysis, and so on. Several notable studies include  Balakrishnan and Chan \cite{Bala}, Raqab and Ahsanullah \cite{rah}, Akhter et al. \cite{ZA3}, Mahmoud et al. \cite{Mah}, Sultan and AL-Thubyani \cite{sal}, Ahsanullah and Alzaatreh  \cite{Ahsan}, Akhter et al. \cite{ZA2} and Guan et al. \cite{Guan}, among others.

\vspace{0.35cm}
Nadarajah \cite{Nad} provided explicit closed form expressions for the moments of order statistics from the normal, log-normal, gamma, and beta distributions. For further insights in this area, one may  refer to Nagaraja \cite{Nag}, Genç \cite{Gen}, MirMostafaee \cite{mir}, Çetinkaya and Genç \cite{Cen}, Akhter et al. \cite{ZA1}, and references therein.

\vspace{0.35cm}
The structure of this paper is laid out as follows: Section 2 outlines the fundamental concepts needed for the upcoming discussions. In Section 3, we present  closed-form expressions for the moments of order statistics. Section 4 builds upon this by offering closed-form expressions for the product moments. Utilizing the results of the previous sections, Section 5 focuses on finding the means, variances, and covariances of order statistics extracted from sample sizes of up to 10 when the shape parameter takes selected values. Section 6 delves into the derivation of Best Linear Unbiased Estimators (BLUEs) and Best Linear Invariant Estimators (BLIEs) for the location and scale parameters of the XLindley distribution, based on both full and Type-II censored data. Moving forward, Section 7 discusses two linear predictors, whereas Sections 8 and 9 provide a simulation study and an example, respectively, for comparative analysis and demonstration. The paper concludes with a summary of key insights in the final section.

\section{Preliminaries}
Suppose $X_{1},\cdots,{X_{n}}$ denote a random sample of size $n$ from the XLindley distribution with pdf $\mathfrak{f}(x)$ and cdf $\digamma(x)$, given in (\ref{pdfxl}) and (\ref{cdfxl}), respectively, and let $X_{1:n}\leq\cdots\leq{X_{n:n}}$ denote the order statistics extracted from the mentioned sample. Then, the pdf of the $r$th order statistic $X_{r:n}$, say $\mathfrak{f}_{r:n}(x)$, for $1\!\leq{r}\!\leq{n}$, is (David and Nagaraja, \cite{david}; Arnold et al., \cite{arnd}) 

\vspace{-0.3cm}
\begin{equation*}
\mathfrak{f}_{r:n}(x)=C_{r:n}(\digamma(x))^{r-1}\{1-{\digamma}(x)\}^{n-r}\mathfrak{f}(x),~~0<x<\infty
\end{equation*}
The joint pdf of $X_{r:n}$ and $X_{s:n}$ is given by
\begin{equation}\label{eqs1}
\mathfrak{f}_{r,s:n}(x,y)=C_{r,s:n}(\digamma(x))^{r-1}\{\digamma(y)-\digamma(x)\}^{s-r-1}\{1-{\digamma}(y)\}^{n-s}\mathfrak{f}(x)\mathfrak{f}(y), 
\end{equation}
where $x$ is less than $y$  and 
\begin{equation}\label{crn}
C_{r:n}=\frac{n!}{(r-1)!(n-r)!}\quad\text{and}\quad C_{r,s:n}=\frac{n!}{(r-1)!(s-r-1)!(n-s)!}.
\end{equation}

Next, the $k$th single moment of $X_{r:n}$ takes the form
\begin{equation}\label{smo}
\varphi^{(k)}_{r:n}=\text{E}{(X^{k}_{r:n})}=\int_{0}^{\infty} x^k \mathfrak{f}_{r:n}(x) {\rm d}x;~~1\leq r\leq n;~ k\in\mathbb{N},
\end{equation}
and the $(k,l)$th product moment of $X_{r:n}$ and $X_{s:n}$ is given by
\begin{equation}\label{dmo}
\varphi^{(k,l)}_{r,s:n}=\text{E}{(X^{k}_{r:n}X^{^l}_{s:n})}=\int_{0}^{\infty}\!\!\int_{x}^{\infty} x^k y^l \mathfrak{f}_{r,s:n}(x,y) {\rm d}y {\rm d}x;~~1\leq r<s\leq n;~ k,l\in\mathbb{N}.
\end{equation}

\section{Single Moments}
The following theorem establishes the main result of this section.
\vspace{0.3cm}

{\bf Theorem~3.1:}~For $1\leq r\leq n$ and $k\in\mathbb{N}$, we have
\begin{equation}\label{T3.1}
\hspace{-0.3in}\varphi^{(k)}_{r:n} = C_{r:n}\sum_{\epsilon =0}^{r-1}\sum_{\rho =0}^{n-r+\epsilon}
\dfrac{\binom{r-1}{\epsilon} \binom{n-r+\epsilon}{\rho}(-1)^\epsilon \psi ^{1-k}\Gamma(\rho + k +1)}{(n-r+\epsilon +1)^{\rho +k+1}(1+\psi)^{2(\rho +1)}}  \left[\psi +2+
\dfrac{\rho + k +1}{\psi (n-r+\epsilon +1)}\right],
\end{equation}
where $\Gamma(.)$ is complete gamma function and  $C_{r:n}$ is given (\ref{crn}).
\vspace{0.3cm}

{\bf Proof:}~~From (\ref{smo}), we can write
\begin{eqnarray*}\nonumber
	\varphi^{(k)}_{r:n}&=&C_{r:n}\int_{0}^{\infty}\!\! x^k [1-(1-\digamma(x))]^{r-1} [1-\digamma(x)]^{n-r} \mathfrak{f}(x) {\rm d}x\\
	&=&\label{pr1}C_{r:n}\sum_{\epsilon =0}^{r-1}(-1)^\epsilon\binom{r-1}{\epsilon}\int_{0}^{\infty}\!\! x^k[1-\digamma(x)]^{n-r+\epsilon} \mathfrak{f}(x)\,{\rm d}x
	\\&=&
	C_{r:n}\sum_{\epsilon =0}^{r-1}(-1)^\epsilon\binom{r\!-\!1}{\epsilon} \int_{0}^{\infty}\!\! 
	x^k\, {\dfrac{\psi^2(2\!+\!\psi\!+\!x)}{(1+\psi)^2}}\bigg(1+\dfrac{\psi x}{(1+\psi)^2}\bigg)^{n-r+\epsilon} {\rm e}^{-\psi(n-r+\epsilon +1)x}\,{\rm d}x
	\\&=&
	C_{r:n}\sum_{\epsilon =0}^{r-1}\sum_{\rho =0}^{n-r+\epsilon}\!(-1)^\epsilon\binom{r\!-\!1}{\epsilon} \binom{n\!-\!r\!+\!\epsilon}{\rho}
	\frac{\psi ^{2+\rho}}{(1+\psi)^{2(\rho +1)}} \int_{0}^{\infty} x^{\rho +k} (2\!+\!\psi\!+\!x)\,{\rm e}^{-\psi(n-r+\epsilon +1)x}\,{\rm d}x.
	\\
\end{eqnarray*}
 
 We note that $\int_0^\infty x^{\nu-1}{\rm e}^{-\eta x}{\rm d}x=\frac{\Gamma(\nu)}{\eta^\nu}$ for $\eta>0$ and $\nu>0,$ see Gradshteyn and Ryzhik \cite{GR},
 so we get
\begin{eqnarray*}
	\varphi^{(k)}_{r:n}&=&
	C_{r:n}\sum_{\epsilon =0}^{r-1}\sum_{\rho =0}^{n-r+\epsilon}\binom{r-1}{\epsilon} \binom{n-r+\epsilon}{\rho}(-1)^\epsilon\!\! 
	\frac{\psi ^{1-k}}{(1+\psi)^{2(\rho +1)}}\\&&\times \left[\dfrac{(\psi +2)\Gamma(\rho + k +1)}{(n-r+\epsilon +1)^{\rho +k+1}}+
	\dfrac{\Gamma(\rho + k +2)}{\psi (n-r+\epsilon +1)^{\rho +k+2}}\right]
	\\&=&
	C_{r:n}\sum_{\epsilon =0}^{r-1}\sum_{\rho =0}^{n-r+\epsilon}
	\dfrac{\binom{r-1}{\epsilon} \binom{n-r+\epsilon}{\rho}(-1)^\epsilon \psi ^{1-k}\Gamma(\rho + k +1)}{(n-r+\epsilon +1)^{\rho +k+1}(1+\psi)^{2(\rho +1)}}  \left[\psi +2+
	\dfrac{\rho + k +1}{\psi (n-r+\epsilon +1)}\right].
\end{eqnarray*}

Thus we have the result. \hfill{$\Box$}\\

{\bf Remark 3.1:}~$\bf(a)$~ By setting $n=r=1$ in (\ref{T3.1}), we obtain
\begin{equation}\label{moment}
\varphi_{1:1}^{(k)}=\text{E}[X^k]=\frac{\psi^2 + 2 \psi +k+1}{\psi^k (1+\psi)^2},
\end{equation}
which is the $k$th moment of $\text{X}$ reported by Chouia and Zeghdoudi \cite{Chouia}.\\


\vspace{0.35cm}
$\bf(b)$~ Upon setting $r=1$ in (\ref{T3.1}),  we have
\begin{equation}
\varphi_{1:n}^{(k)}=\, \sum_{\rho=0}^{n-1}\,\binom{n-1}{\rho} \frac{\psi^{1-k}\Gamma(\rho+k+2)}{n^{\rho+k}(1+\psi)^{2(\rho+1)}} \left( \psi +2 + \frac{\rho +k+1}{n \psi} \right),
\end{equation}
and for $r=n$, in (\ref{T3.1}), we have
\begin{equation}
\varphi_{n:n}^{(k)}=n\,  \sum_{\epsilon=0}^{n -1}\,\sum_{\rho =0}^{\epsilon} \, 
(-1)^{\epsilon} \binom{n -1}{\epsilon} \, \binom{\epsilon}{\rho}
 \frac{\psi^{1-k} \Gamma(\rho+k+2)}{(1+\psi)^{2(\rho+1)} (\epsilon+1)^{\rho +k+1}} \left( \psi +2 + \frac{\rho +k+1}{\psi(\epsilon+1)} \right),
\end{equation}
which are the $k$th moments of the minimum and the maximum order statistics, respectively.

\vspace{.2cm}
 \section{Product Moments}
In this section, we obtain the results for the product moments of order statistics from the XLindley distribution in the following theorem.\\

{\bf Theorem~4.1:}~For $1\leq r<s\leq n$, $k,l\in\mathbb{N}$, we obtain the closed-form expression for product moments of order statistics;
\begin{eqnarray*} 
	\varphi^{(k,l)}_{r,s:n}&=&C_{r,s:n}\sum_{\epsilon=0}^{r-1}\sum_{\rho=0}^{s-r-1}\sum_{\textgamma=0}^{n-r-1-\rho}
	\sum_{\eta=0}^{\epsilon+\rho}\binom{r\!-\!1}{\epsilon}\binom{s-r-1}{\rho}\binom{n-r-1-\rho}{\textgamma}\binom{\epsilon+\rho}{\eta}
	\\&&\times 
	\dfrac{(l+\textgamma)! (-1)^{\epsilon+s-r-1-\rho}}{(n-r-\rho) (n-r+\epsilon+1)^{k+\eta+1} (1+\psi)^{2(2+\eta+\textgamma)}\psi^{k+l-2}  } \\&&\times
	\bigg[\frac{\big(\psi+2 +\frac{k+\eta+l+\textgamma+2}{\psi(n-r+\epsilon+1)}\big) \Gamma(k+\eta+l+\textgamma+2)}{\psi(n-r+\epsilon+1)^{l+\textgamma+1}(l+\textgamma)!}\\
\end{eqnarray*}
\begin{eqnarray}\label{4.1}
	&&\quad +\left(\psi +2+\frac{l+\textgamma+1}{\psi (n-r-\rho)}\right)\sum_{\delta=0}^{l+\textgamma} \frac{  \big(\psi+2+
		\frac{k+\eta+\delta+1}{\psi(n-r+\epsilon+1)}\big) 
		\Gamma(k+\eta+\delta+1)}{(n-r-\rho)^{l+\textgamma -\delta }\, (n-r+\epsilon+1)^{\delta}\, \delta !}
	\bigg],
\end{eqnarray}

where $C_{r,s:n}$ is defined as before.
\vspace{0.5cm}

{\bf Proof:}~~ From (\ref{eqs1}), we have
\vspace{.2cm}
\begin{eqnarray}\label{p4-11}\nonumber
\varphi^{(k,l)}_{r,s:n}&=&C_{r,s:n}\!\!\int_{0}^{\infty}\!\!\!\int_{x}^{\infty}\!\!\!x^{k}y^{l}{[1-(1-\digamma(x))]}^{r-1}{[1\!-\!\digamma(y)]}^{n-s}{[(1-\digamma(x))\!-\!(1-\digamma(y))]}
^{s-r-1}
\\&&\quad\quad\quad\quad \quad\quad\quad\quad \times {\mathfrak{f}(x)}{\mathfrak{f}(y)}{{\rm d}y}{{\rm d}x}.
\end{eqnarray}

Utilizing the binomial expansion, (\ref{p4-11}) can be reexpressed as,

\begin{equation*}\label{p4-2}
\varphi^{(k,l)}_{r,s:n}=~C_{r,s:n}\!\!\sum_{\epsilon=0}^{r-1}\sum_{\rho=0}^{s-r-1}(-1)^{\epsilon+s-r-1-\rho}\binom{r\!-\!1}{\epsilon}\binom{s-r-1}{\rho}
\!\!\int_{0}^{\infty}\!\!\!x^{k}{[1-\digamma(x)]}^{\epsilon +\rho}\mathfrak{f}(x)\text{I}(x){\rm d}x,
\end{equation*}
where
\begin{eqnarray*}
	\text{I}(x)&=&\int_{x}^{\infty}\!\!y^{l}{[1-\digamma(y)]}^{n-r-1-\rho}{\mathfrak{f}(y)}{{\rm d}y}\\&=&
	\dfrac{\psi^2}{(1+\psi)^2}\int_{x}^{\infty}\!\!y^{l}\bigg(1+\frac{\psi y}{(1+\psi)^2}\bigg)^{n-r-1-\rho}(\psi +2+y){\rm e}^{-\psi (n-r-\rho)y}{\rm d}y
	\\&=&
	\sum_{\textgamma=0}^{n-r-1-\rho}\binom{n-r-1-\rho}{\textgamma}\dfrac{\psi^{2+\textgamma}}{(1+\psi)^{2(1+\textgamma)}}
	\int_{x}^{\infty}\!\!y^{l+\textgamma}(\psi +2+y){\rm e}^{-\psi (n-r-\rho)y}{\rm d}y \\
\end{eqnarray*}
Note that $\int_u^\infty x^{\nu}{\rm e}^{-\eta x}{\rm d}x={\rm e}^{-\eta u}\sum_{p=0}^\nu\frac{\nu!}{p!}\frac{u^p}{\eta^{\nu-p+1}}$
for $\eta >0~u>0,$ and $\nu =0,1,2,\dots,$  see Gradshteyn and Ryzhik \cite{GR}, so we get
\begin{eqnarray*} 
		\text{I}(x) &=&
	\sum_{\textgamma=0}^{n-r-1-\rho}\binom{n-r-1-\rho}{\textgamma}\dfrac{\psi^{2+\textgamma}}{(1+\psi)^{2(1+\textgamma)}}\, {\rm e}^{-\psi (n-r-\rho)x} \\&&\times
	\left[(\psi +2)\sum_{\delta=0}^{l+\textgamma} \frac{x^\delta (l+\textgamma)!}{(\psi (n-r-\rho))^{l+\textgamma -\delta +1}\delta !} +
	\sum_{\delta=0}^{l+\textgamma+1} \frac{x^\delta (l+\textgamma+1)!}{(\psi (n-r-\rho))^{l+\textgamma -\delta +2}\delta !}\right]
	\\&=&
	\sum_{\textgamma=0}^{n-r-1-\rho}\binom{n-r-1-\rho}{\textgamma}\dfrac{\psi^{2+\textgamma} (l+\textgamma)!}{(1+\psi)^{2(1+\textgamma)}}\, {\rm e}^{-\psi (n-r-\rho)x} \\&&\times
	\left[(\psi +2)\sum_{\delta=0}^{l+\textgamma} \frac{x^\delta }{(\psi (n-r-\rho))^{l+\textgamma -\delta +1}\delta !} +
	\sum_{\delta=0}^{l+\textgamma+1} \frac{x^\delta (l+\textgamma+1)}{(\psi (n-r-\rho))^{l+\textgamma -\delta +2}\delta !}\right]
	\\&=&
	\sum_{\textgamma=0}^{n-r-1-\rho}\binom{n-r-1-\rho}{\textgamma}\dfrac{\psi^{1+\textgamma} (l+\textgamma)! {\rm e}^{-\psi (n-r-\rho)x}}{(1+\psi)^{2(1+\textgamma)}(n-r-\rho)} 
	\left[\frac{x^{l+\textgamma+1}}{(l+\textgamma)!  }+\sum_{\delta=0}^{l+\textgamma} \frac{ \left(\psi +2+\frac{l+\textgamma+1}{\psi (n-r-\rho)}\right)\, x^\delta}{[\psi (n-r-\rho)]^{l+\textgamma -\delta }\delta !}
	\right].
\end{eqnarray*}
Therefore, we have
\begin{eqnarray*}
	&&\int_{0}^{\infty}\!\!\!x^{k}{[1-\digamma(x)]}^{\epsilon +\rho}\mathfrak{f}(x)\text{I}(x){\rm d}x=
	\sum_{\textgamma=0}^{n-r-1-\rho}\binom{n-r-1-\rho}{\textgamma}\dfrac{\psi^{1+\textgamma} (l+\textgamma)! }{(1+\psi)^{2(1+\textgamma)}(n-r-\rho)} 
	\\&&\times \int_0^\infty x^{k}{[1-\digamma(x)]}^{\epsilon +\rho}\mathfrak{f}(x) {\rm e}^{-\psi (n-r-\rho)x}
	\left[\frac{x^{l+\textgamma+1}}{(l+\textgamma)!  }+\sum_{\delta=0}^{l+\textgamma} \frac{ \left(\psi +2+\frac{l+\textgamma+1}{\psi (n-r-\rho)}\right)\, x^\delta}{[\psi (n-r-\rho)]^{l+\textgamma -\delta }\delta !} 
	\right]{\rm d}x.
\end{eqnarray*}
Applying the result $\quad\int_0^\infty x^{\nu-1}{\rm e}^{-\eta x}{\rm d}x=\frac{\Gamma(\nu)}{\eta^\nu}$ for $ \eta>0$ and $\nu>0,$  (see  Gradshteyn and Ryzhik \cite{GR}), we obtain

\begin{eqnarray*}
	&& \int_0^\infty x^{k}{[1-\digamma(x)]}^{\epsilon +\rho}\mathfrak{f}(x) {\rm e}^{-\psi (n-r-\rho)x}
	\left[\frac{x^{l+\textgamma+1}}{(l+\textgamma)!  }+\sum_{\delta=0}^{l+\textgamma} \frac{ \left(\psi +2+\frac{l+\textgamma+1}{\psi (n-r-\rho)}\right)\, x^\delta}{[\psi (n-r-\rho)]^{l+\textgamma -\delta }\delta !} 
	\right]{\rm d}x
	\\&=&
	\frac{\psi^2}{(1+\psi)^2}\int_0^\infty x^{k}\left(1+\frac{\psi x}{(1+\psi)^2}\right)^{\epsilon +\rho} (\psi +2+x){\rm e}^{-\psi (n-r+\epsilon +1)x}
	\\&&\times\left[\frac{x^{l+\textgamma+1}}{(l+\textgamma)!  }+\sum_{\delta=0}^{l+\textgamma} \frac{ \left(\psi +2+\frac{l+\textgamma+1}{\psi (n-r-\rho)}\right)\, x^\delta}{[\psi (n-r-\rho)]^{l+\textgamma -\delta }\delta !} 
	\right]{\rm d}x
	\\&=&\sum_{\eta=0}^{\epsilon+\rho}\binom{\epsilon+\rho}{\eta}
	\frac{\psi^{2+\eta}}{(1+\psi)^{2(1+\eta)}}\int_0^\infty x^{k+\eta} (\psi +2+x){\rm e}^{-\psi (n-r+\epsilon +1)x}
	\\&&\times\left[\frac{x^{l+\textgamma+1}}{(l+\textgamma)!  }+\sum_{\delta=0}^{l+\textgamma} \frac{ \left(\psi +2+\frac{l+\textgamma+1}{\psi (n-r-\rho)}\right)\, x^\delta}{[\psi (n-r-\rho)]^{l+\textgamma -\delta }\delta !} 
	\right]{\rm d}x
	\\&=&\sum_{\eta=0}^{\epsilon+\rho}\binom{\epsilon+\rho}{\eta}
	\frac{\psi^{2+\eta}}{(1+\psi)^{2(1+\eta)}}\bigg[\frac{\big(\psi+2 +\frac{k+\eta+l+\textgamma+2}{\psi(n-r+\epsilon+1)}\big) \Gamma(k+\eta+l+\textgamma+2)}{[\psi(n-r+\epsilon+1)]^{k+\eta+l+\textgamma+2}(l+\textgamma)!}
	\\&&+\left(\psi +2+\frac{l+\textgamma+1}{\psi (n-r-\rho)}\right)\sum_{\delta=0}^{l+\textgamma} \frac{  \big(\psi+2+
		\frac{k+\eta+\delta+1}{\psi(n-r+\epsilon+1)}\big) 
		\Gamma(k+\eta+\delta+1)}{[\psi (n-r-\rho)]^{l+\textgamma -\delta }\delta !\, [\psi(n-r+\epsilon+1)]^{k+\eta+\delta+1}}
	\bigg].
\end{eqnarray*}
Finally, we can write
\begin{eqnarray*}
	\varphi^{(k,l)}_{r,s:n}&=&C_{r,s:n}\sum_{\epsilon=0}^{r-1}\sum_{\rho=0}^{s-r-1}\sum_{\textgamma=0}^{n-r-1-\rho}
	\sum_{\eta=0}^{\epsilon+\rho}\binom{r\!-\!1}{\epsilon}\binom{s-r-1}{\rho}\binom{n-r-1-\rho}{\textgamma}\binom{\epsilon+\rho}{\eta}
	\\&&\times (-1)^{\epsilon+s-r-1-\rho}
	\dfrac{\psi^{1+\textgamma} (l+\textgamma)! }{(1+\psi)^{2(1+\textgamma)}(n-r-\rho)} 
	\frac{\psi^{2+\eta}}{(1+\psi)^{2(1+\eta)}}
	\\&&\times \bigg[\frac{\big(\psi+2 +\frac{k+\eta+l+\textgamma+2}{\psi(n-r+\epsilon+1)}\big) \Gamma(k+\eta+l+\textgamma+2)}{[\psi(n-r+\epsilon+1)]^{k+\eta+l+\textgamma+2}(l+\textgamma)!}
	\\&&\quad +\left(\psi +2+\frac{l+\textgamma+1}{\psi (n-r-\rho)}\right)\sum_{\delta=0}^{l+\textgamma} \frac{  \big(\psi+2+
		\frac{k+\eta+\delta+1}{\psi(n-r+\epsilon+1)}\big) 
		\Gamma(k+\eta+\delta+1)}{[\psi (n-r-\rho)]^{l+\textgamma -\delta }\delta !\, [\psi(n-r+\epsilon+1)]^{k+\eta+\delta+1}}
	\bigg]
	\\&=&C_{r,s:n}\sum_{\epsilon=0}^{r-1}\sum_{\rho=0}^{s-r-1}\sum_{\textgamma=0}^{n-r-1-\rho}
	\sum_{\eta=0}^{\epsilon+\rho}\binom{r\!-\!1}{\epsilon}\binom{s-r-1}{\rho}\binom{n-r-1-\rho}{\textgamma}\binom{\epsilon+\rho}{\eta}
	\\&&\times 
	\dfrac{\psi^{3+\textgamma+\eta} (l+\textgamma)! (-1)^{\epsilon+s-r-1-\rho}}{(1+\psi)^{2(2+\eta+\textgamma)}(n-r-\rho)} \\&&\times
	\bigg[\frac{\big(\psi+2 +\frac{k+\eta+l+\textgamma+2}{\psi(n-r+\epsilon+1)}\big) \Gamma(k+\eta+l+\textgamma+2)}{[\psi(n-r+\epsilon+1)]^{k+\eta+l+\textgamma+2}(l+\textgamma)!}
	\\&&\quad +\left(\psi +2+\frac{l+\textgamma+1}{\psi (n-r-\rho)}\right)\sum_{\delta=0}^{l+\textgamma} \frac{  \big(\psi+2+
		\frac{k+\eta+\delta+1}{\psi(n-r+\epsilon+1)}\big) 
		\Gamma(k+\eta+\delta+1)}{[\psi (n-r-\rho)]^{l+\textgamma -\delta }\delta !\, [\psi(n-r+\epsilon+1)]^{k+\eta+\delta+1}}
	\bigg].
\end{eqnarray*}
Hence, the result (\ref{4.1}) can be obtained after some algebraic manipulation. \hfill{$\Box$}

\vspace{0.30 cm}
\section{Computations of Means, Variances and Covariances}
\vspace{-0.1cm}
Here, we build upon the results derived in the preceding sections to calculate the means, variances, and covariances of order statistics for the XLindley distribution.

\vspace{0.2cm}
To find the means of the order statistics, we substitute $k=1$ into (\ref{T3.1}). This allows us to calculate the means for sample sizes ranging from  $n = 1$ to $n=10$ and for  selected values of the shape parameter. Table 1 includes the computed mean values.

\vspace{0.2cm}
\setlength{\tabcolsep}{2.1em}
\small{\bf Table 1:} Means of order statistics  \vspace{-0.5cm}
\begin{center}
\begin{longtable}{|c|c|c|c|c|c|p{2cm}|p{2cm}|p{2cm}|p{2cm}|}
\multicolumn{6}{r}{({\em Continued})}\\[.5ex]
\endfoot
\endlastfoot
\hline
$n$ & $r$ & $~~~\psi=1$ & $~~~~\psi=2$ & $~~~\psi=3$ & $~~~~\psi=4$\\
\hline
$1$	&	$1$	&	$1.25000$	&	$0.55556$	&	$0.35417$	&	$0.26000$	\\

$2$	&	$1$	&	$0.64062$	&	$0.27932$	&	$0.17741$	&	$0.13010$	\\
	&	$2$	&	$1.85938$	&	$0.83179$	&	$0.53092$	&	$0.38990$	\\
	
$3$	&	$1$	&	$0.43171$	&	$0.18661$	&	$0.11835$	&	$0.08676$	\\
	&	$2$	&	$1.05845$	&	$0.46475$	&	$0.29552$	&	$0.21679$	\\
	&	$3$	&	$2.25984$	&	$1.01531$	&	$0.64862$	&	$0.47646$	\\

$4$	&	$1$	&	$0.32578$	&	$0.14011$	&	$0.08879$	&	$0.06508$	\\
	&	$2$	&	$0.74953$	&	$0.32609$	&	$0.20702$	&	$0.15180$	\\
	&	$3$	&	$1.36737$	&	$0.60340$	&	$0.38403$	&	$0.28178$	\\
	&	$4$	&	$2.55733$	&	$1.15261$	&	$0.73682$	&	$0.54135$	\\

$5$	&	$1$	&	$0.26166$	&	$0.11217$	&	$0.07105$	&	$0.05207$	\\
	&	$2$	&	$0.58225$	&	$0.25189$	&	$0.15977$	&	$0.11712$	\\
	&	$3$	&	$1.00045$	&	$0.43740$	&	$0.27790$	&	$0.20381$	\\
	&	$4$	&	$1.61199$	&	$0.71407$	&	$0.45477$	&	$0.33375$	\\
	&	$5$	&	$2.79366$	&	$1.26225$	&	$0.80734$	&	$0.59325$	\\

$6$	&	$1$	&	$0.21866$	&	$0.09352$	&	$0.05922$	&	$0.04339$	\\
	&	$2$	&	$0.47666$	&	$0.20542$	&	$0.13022$	&	$0.09544$	\\
	&	$3$	&	$0.79342$	&	$0.34483$	&	$0.21888$	&	$0.16048$	\\
	&	$4$	&	$1.20748$	&	$0.52996$	&	$0.33693$	&	$0.24715$	\\
	&	$5$	&	$1.81424$	&	$0.80613$	&	$0.51370$	&	$0.37705$	\\
	&	$6$	&	$2.98954$	&	$1.35348$	&	$0.86606$	&	$0.63649$	\\

$7$	&	$1$	&	$0.18781$	&	$0.08018$	&	$0.05076$	&	$0.03719$	\\
	&	$2$	&	$0.40375$	&	$0.17351$	&	$0.10994$	&	$0.08057$	\\
	&	$3$	&	$0.65894$	&	$0.28519$	&	$0.18090$	&	$0.13261$	\\
	&	$4$	&	$0.97273$	&	$0.42435$	&	$0.26951$	&	$0.19763$	\\
	&	$5$	&	$1.38354$	&	$0.60916$	&	$0.38750$	&	$0.28428$	\\
	&	$6$	&	$1.98653$	&	$0.88491$	&	$0.56417$	&	$0.41416$	\\
	&	$7$	&	$3.15671$	&	$1.43157$	&	$0.91638$	&	$0.67355$	\\

$8$	&	$1$	&	$0.16460$	&	$0.07018$	&	$0.04442$	&	$0.03255$	\\
	&	$2$	&	$0.35030$	&	$0.15022$	&	$0.09516$	&	$0.06973$	\\
	&	$3$	&	$0.56408$	&	$0.24338$	&	$0.15430$	&	$0.11310$	\\
	&	$4$	&	$0.81703$	&	$0.35488$	&	$0.22523$	&	$0.16513$	\\
	&	$5$	&	$1.12843$	&	$0.49382$	&	$0.31379$	&	$0.23014$	\\
	&	$6$	&	$1.53660$	&	$0.67837$	&	$0.43173$	&	$0.31677$	\\
	&	$7$	&	$2.13650$	&	$0.95376$	&	$0.60832$	&	$0.44662$	\\
	&	$8$	&	$3.30246$	&	$1.49983$	&	$0.96039$	&	$0.70596$	\\

$9$	&	$1$	&	$0.14649$	&	$0.06239$	&	$0.03949$	&	$0.02893$	\\
	&	$2$	&	$0.30941$	&	$0.13246$	&	$0.08389$	&	$0.06147$	\\
	&	$3$	&	$0.49341$	&	$0.21237$	&	$0.13460$	&	$0.09865$	\\
	&	$4$	&	$0.70543$	&	$0.30540$	&	$0.19372$	&	$0.14201$	\\
	&	$5$	&	$0.95653$	&	$0.41674$	&	$0.26461$	&	$0.19403$	\\
	&	$6$	&	$1.26595$	&	$.555490$	&	$0.35314$	&	$0.25903$	\\
	&	$7$	&	$1.67193$	&	$0.73981$	&	$0.47102$	&	$0.34564$	\\
	&	$8$	&	$2.26924$	&	$1.01489$	&	$0.64755$	&	$0.47547$	\\
	&	$9$	&	$3.43161$	&	$1.56045$	&	$0.99949$	&	$0.73477$	\\

$10$&	$1$	&	$0.13198$	&	$0.05616$	&	$0.03554$	&	$0.02604$	\\
	&	$2$	&	$0.27711$	&	$0.11847$	&	$0.07501$	&	$0.05496$	\\
	&	$3$	&	$0.43864$	&	$0.18843$	&	$0.11939$	&	$0.08750$	\\
	&	$4$	&	$0.62121$	&	$0.26824$	&	$0.17008$	&	$0.12467$	\\
	&	$5$	&	$0.83175$	&	$0.36114$	&	$0.22918$	&	$0.16802$	\\
	&	$6$	&	$1.08130$	&	$0.47234$	&	$0.30004$	&	$0.22003$	\\
	&	$7$	&	$1.38905$	&	$0.61093$	&	$0.38853$	&	$0.28502$	\\
	&	$8$	&	$1.79317$	&	$0.79504$	&	$0.50638$	&	$0.37162$	\\
	&	$9$	&	$2.38825$	&	$1.06985$	&	$0.68284$	&	$0.50144$	\\
	&	$10$&	$3.54754$	&	$1.61496$	&	$1.03468$	&	$0.76070$	\\
\hline
\end{longtable}
\end{center}

\vspace{-0.9cm}
We see that the condition $\sum_{r=1}^n \varphi_{r:n}=n\text{E}(X)$, is satisfied (David and Nagaraja \cite{david})

\vspace{0.5cm}
By setting $k=1$ and $k=2$ in (\ref{T3.1}), respectively, we can systematically calculate the first two moments $\varphi_{r:n}^{(1)}$ and $\varphi_{r:n}^{(2)}$.
 These moments facilitate the computation of the variances of the order statistics. We have compiled tables of variances for sample sizes ranging from  $n = 1$ to $n=10$ and for  selected values of the shape parameter.

\vspace{0.3cm}
By setting $k=l=1$ in (\ref{4.1}), we first compute all product moments. Utilizing these product moments in conjunction with the previously calculated first two moments allows for straightforward computation of the covariances of order statistics.  We calculated the covariances for for sample sizes ranging from  $n = 1$ to $n=10$ and for  selected values of the shape parameter. The resulting variances and covariances are reported in Table 2.

\vspace{0.3cm}
\setlength{\tabcolsep}{1.53em}
{\bf Table 2:} Variances and covariances of order statistics.\vspace{-0.2cm}
\small
\begin{longtable}{|c|c|c|p{1.7cm}|p{1.7cm}|p{1.7cm}|p{1.7cm}|}
\multicolumn{7}{r}{{({\em Continued})}}\\[.7ex]
\endfoot
\endlastfoot
\hline
$n$ & $s$ & $r$ & $~~~\psi=1$ & $~~~\psi=2$ & $~~\psi=3$ & $~~~~\psi=4$\\
\hline
$1$	&	$1$ &	$1$	& 	$~~1.43750$ & 	$~~0.30247$	&	$~~0.12457$ & 	$~~0.06740$	\\
$2$	&	$1$ &	$1$	& 	$~~0.38647$ &	$~~0.07707$	&	$~~0.03135$ &	$~~0.01690$	\\
	&	$2$ &	$1$	& 	$~~0.37134$ &	$~~0.07631$	&	$~~0.03124$ &	$~~0.01687$	\\
	&	$2$	&	$2$	& 	$~~1.74585$ &	$~~0.37525$	&	$~~0.15529$ &	$~~0.08415$	\\
$3$	&	$1$	&	$1$	& 	$~~0.17782$ &	$~~0.03452$	&	$~~0.01397$ &	$~~0.00752$	\\
	&	$2$	&	$1$	& 	$~~0.17302$ &	$~~0.03432$	&	$~~0.01394$ &	$~~0.00751$	\\
	&	$3$	&	$1$	& 	$~~0.16681$ &	$~~0.03399$	&	$~~0.01390$ &	$~~0.00750$	\\
	&	$2$	&	$2$	& 	$~~0.54192$ &	$~~0.11061$	&	$~~0.04519$ &	$~~0.02439$	\\
	&	$3$	&	$2$	&	$~~0.52320$	&	$~~0.10957$	&	$~~0.04504$	&	$~~0.02435$	\\
	&	$3$	&	$3$	&	$~~1.86670$	&	$~~0.40654$	&	$~~0.16879$	&	$~~0.09156$	\\
$4$	&	$1$	&	$1$	&	$~~0.10211$	&	$~~0.01950$	&	$~~0.00787$	&	$~~0.00423$	\\
	&	$2$	&	$1$	&	$~~0.09995$	&	$~~0.01942$	&	$~~0.00786$	&	$~~0.00423$	\\
	&	$3$	&	$1$	&	$~~0.09749$	&	$~~0.01931$	&	$~~0.00784$	&	$~~0.00423$	\\
	&	$4$	&	$1$	&	$~~0.09425$	&	$~~0.01913$	&	$~~0.00782$	&	$~~0.00422$	\\
	&	$2$	&	$2$	&	$~~0.27029$	&	$~~0.05364$	&	$~~0.02179$	&	$~~0.01174$	\\
	&	$3$	&	$2$	&	$~~0.26379$	&	$~~0.05334$	&	$~~0.02175$	&	$~~0.01173$	\\
	&	$4$	&	$2$	&	$~~0.25518$	&	$~~0.05285$	&	$~~0.02168$	&	$~~0.01171$	\\
	&	$3$	&	$3$	&	$~~0.62267$	&	$~~0.12912$	&	$~~0.05293$	&	$~~0.02859$	\\
	&	$4$	&	$3$	&	$~~0.60312$	&	$~~0.12797$	&	$~~0.05275$	&	$~~0.02855$	\\
	&	$4$	&	$4$	&	$~~1.92738$	&	$~~0.42360$	&	$~~0.17629$	&	$~~0.09570$	\\
$5$	&	$1$	&	$1$	&	$~~0.06625$	&	$~~0.01251$	&	$~~0.00504$	&	$~~0.00271$	\\
	&	$2$	&	$1$	&	$~~0.06509$	&	$~~0.01247$	&	$~~0.00503$	&	$~~0.00271$	\\
	&	$3$	&	$1$	&	$~~0.06384$	&	$~~0.01242$	&	$~~0.00503$	&	$~~0.00271$	\\
	&	$4$	&	$1$	&	$~~0.06238$	&	$~~0.01235$	&	$~~0.00502$	&	$~~0.00270$	\\
	&	$5$	&	$1$	&	$~~0.06044$	&	$~~0.01224$	&	$~~0.00500$	&	$~~0.00270$	\\
	&	$2$	&	$2$	&	$~~0.16330$	&	$~~0.03182$	&	$~~0.01288$	&	$~~0.00694$	\\
	&	$3$	&	$2$	&	$~~0.16019$	&	$~~0.03170$	&	$~~0.01287$	&	$~~0.00693$	\\
	&	$4$	&	$2$	&	$~~0.15659$	&	$~~0.03153$	&	$~~0.01284$	&	$~~0.00693$	\\
	&	$5$	&	$2$	&	$~~0.15177$	&	$~~0.03124$	&	$~~0.01280$	&	$~~0.00692$	\\
	&	$3$	&	$3$	&	$~~0.32584$	&	$~~0.06571$	&	$~~0.02677$	&	$~~0.01444$	\\
	&	$4$	&	$3$	&	$~~0.31868$	&	$~~0.06536$	&	$~~0.02672$	&	$~~0.01442$	\\
	&	$5$	&	$3$	&	$~~0.30904$	&	$~~0.06478$	&	$~~0.02664$	&	$~~0.01440$	\\
	&	$4$	&	$4$	&	$~~0.67097$	&	$~~0.14078$	&	$~~0.05785$	&	$~~0.03127$	\\
	&	$5$	&	$4$	&	$~~0.65142$	&	$~~0.13956$	&	$~~0.05766$	&	$~~0.03123$	\\
	&	$5$	&	$5$	&	$~~1.96222$	&	$~~0.43421$	&	$~~0.18104$	&	$~~0.09834$	\\
$6$	&	$1$	&	$1$	&	$~~0.04646$	&	$~~0.00870$	&	$~~0.00350$	&	$~~0.00188$	\\
	&	$2$	&	$1$	&	$~~0.04577$	&	$~~0.00868$	&	$~~0.00350$	&	$~~0.00188$	\\
	&	$3$	&	$1$	&	$~~0.04503$	&	$~~0.00865$	&	$~~0.00350$	&	$~~0.00188$	\\
	&	$4$	&	$1$	&	$~~0.04422$	&	$~~0.00862$	&	$~~0.00349$	&	$~~0.00188$	\\
	&	$5$	&	$1$	&	$~~0.04328$	&	$~~0.00857$	&	$~~0.00348$	&	$~~0.00188$	\\
	&	$6$	&	$1$	&	$~~0.04201$	&	$~~0.00850$	&	$~~0.00347$	&	$~~0.00188$	\\
	&	$2$	&	$2$	&	$~~0.10972$	&	$~~0.02111$	&	$~~0.00853$	&	$~~0.00459$	\\
	&	$3$	&	$2$	&	$~~0.10798$	&	$~~0.02104$	&	$~~0.00852$	&	$~~0.00459$	\\
	&	$4$	&	$2$	&	$~~0.10606$	&	$~~0.02096$	&	$~~0.00851$	&	$~~0.00458$	\\
	&	$5$	&	$2$	&	$~~0.10382$	&	$~~0.02085$	&	$~~0.00849$	&	$~~0.00458$	\\
	&	$6$	&	$2$	&	$~~0.10079$	&	$~~0.02067$	&	$~~0.00847$	&	$~~0.00457$	\\
	&	$3$	&	$3$	&	$~~0.20356$	&	$~~0.04029$	&	$~~0.01636$	&	$~~0.00881$	\\
	&	$4$	&	$3$	&	$~~0.20000$	&	$~~0.04014$	&	$~~0.01634$	&	$~~0.00881$	\\
	&	$5$	&	$3$	&	$~~0.19584$	&	$~~0.03993$	&	$~~0.01631$	&	$~~0.00880$	\\
	&	$6$	&	$3$	&	$~~0.19017$	&	$~~0.03958$	&	$~~0.01625$	&	$~~0.00879$	\\
	&	$4$	&	$4$	&	$~~0.36239$	&	$~~0.07399$	&	$~~0.03022$	&	$~~0.01631$	\\
	&	$5$	&	$4$	&	$~~0.35500$	&	$~~0.07361$	&	$~~0.03016$	&	$~~0.01629$	\\
	&	$6$	&	$4$	&	$~~0.34492$	&	$~~0.07298$	&	$~~0.03007$	&	$~~0.01627$	\\
	&	$5$	&	$5$	&	$~~0.70254$	&	$~~0.14875$	&	$~~0.06125$	&	$~~0.03313$	\\
	&	$6$	&	$5$	&	$~~0.68330$	&	$~~0.14750$	&	$~~0.06106$	&	$~~0.03308$	\\
	&	$6$	&	$6$	&	$~~1.98393$	&	$~~0.44137$	&	$~~0.18430$	&	$~~0.10017$	\\
$7$	&	$1$	&	$1$	&	$~~0.03439$	&	$~~0.00640$	&	$~~0.00257$	&	$~~0.00138$	\\
	&	$2$	&	$1$	&	$~~0.03394$	&	$~~0.00639$	&	$~~0.00257$	&	$~~0.00138$	\\
	&	$3$	&	$1$	&	$~~0.03347$	&	$~~0.00637$	&	$~~0.00257$	&	$~~0.00138$	\\
	&	$4$	&	$1$	&	$~~0.03297$	&	$~~0.00635$	&	$~~0.00257$	&	$~~0.00138$	\\
	&	$5$	&	$1$	&	$~~0.03241$	&	$~~0.00633$	&	$~~0.00256$	&	$~~0.00138$	\\
	&	$6$	&	$1$	&	$~~0.03176$	&	$~~0.00629$	&	$~~0.00256$	&	$~~0.00138$	\\
	&	$7$	&	$1$	&	$~~0.03087$	&	$~~0.00624$	&	$~~0.00255$	&	$~~0.00138$	\\
	&	$2$	&	$2$	&	$~~0.07894$	&	$~~0.01504$	&	$~~0.00607$	&	$~~0.00326$	\\
	&	$3$	&	$2$	&	$~~0.07785$	&	$~~0.01500$	&	$~~0.00606$	&	$~~0.00326$	\\
	&	$4$	&	$2$	&	$~~0.07669$	&	$~~0.01496$	&	$~~0.00606$	&	$~~0.00326$	\\
	&	$5$	&	$2$	&	$~~0.07541$	&	$~~0.01490$	&	$~~0.00605$	&	$~~0.00326$	\\
	&	$6$	&	$2$	&	$~~0.07391$	&	$~~0.01482$	&	$~~0.00604$	&	$~~0.00326$	\\
	&	$7$	&	$2$	&	$~~0.07185$	&	$~~0.01470$	&	$~~0.00602$	&	$~~0.00325$	\\
	&	$3$	&	$3$	&	$~~0.14018$	&	$~~0.02737$	&	$~~0.01108$	&	$~~0.00597$	\\
	&	$4$	&	$3$	&	$~~0.13811$	&	$~~0.02729$	&	$~~0.01107$	&	$~~0.00596$	\\
	&	$5$	&	$3$	&	$~~0.13583$	&	$~~0.02719$	&	$~~0.01106$	&	$~~0.00596$	\\
	&	$6$	&	$3$	&	$~~0.13315$	&	$~~0.02705$	&	$~~0.01104$	&	$~~0.00596$	\\
	&	$7$	&	$3$	&	$~~0.12946$	&	$~~0.02682$	&	$~~0.01101$	&	$~~0.00595$	\\
	&	$4$	&	$4$	&	$~~0.23181$	&	$~~0.04645$	&	$~~0.01890$	&	$~~0.01019$	\\
	&	$5$	&	$4$	&	$~~0.22803$	&	$~~0.04628$	&	$~~0.01888$	&	$~~0.01018$	\\
	&	$6$	&	$4$	&	$~~0.22357$	&	$~~0.04604$	&	$~~0.01884$	&	$~~0.01017$	\\
	&	$7$	&	$4$	&	$~~0.21744$	&	$~~0.04566$	&	$~~0.01878$	&	$~~0.01016$	\\
	&	$5$	&	$5$	&	$~~0.38800$	&	$~~0.08001$	&	$~~0.03274$	&	$~~0.01768$	\\
	&	$6$	&	$5$	&	$~~0.38057$	&	$~~0.07961$	&	$~~0.03268$	&	$~~0.01766$	\\
	&	$7$	&	$5$	&	$~~0.37032$	&	$~~0.07895$	&	$~~0.03258$	&	$~~0.01764$	\\
	&	$6$	&	$6$	&	$~~0.72447$	&	$~~0.15452$	&	$~~0.06373$	&	$~~0.03449$	\\
	&	$7$	&	$6$	&	$~~0.70564$	&	$~~0.15326$	&	$~~0.06354$	&	$~~0.03445$	\\
	&	$7$	&	$7$	&	$~~1.99822$	&	$~~0.44648$	&	$~~0.18668$	&	$~~0.10150$	\\
$8$	&	$1$	&	$1$	&	$~~0.02648$	&	$~~0.00491$	&	$~~0.00197$	&	$~~0.00106$	\\
	&	$2$	&	$1$	&	$~~0.02617$	&	$~~0.00490$	&	$~~0.00197$	&	$~~0.00106$	\\
	&	$3$	&	$1$	&	$~~0.02585$	&	$~~0.00489$	&	$~~0.00197$	&	$~~0.00106$	\\
	&	$4$	&	$1$	&	$~~0.02552$	&	$~~0.00487$	&	$~~0.00197$	&	$~~0.00106$	\\
	&	$5$	&	$1$	&	$~~0.02516$	&	$~~0.00486$	&	$~~0.00197$	&	$~~0.00106$	\\
	&	$6$	&	$1$	&	$~~0.02476$	&	$~~0.00484$	&	$~~0.00196$	&	$~~0.00106$	\\
	&	$7$	&	$1$	&	$~~0.02429$	&	$~~0.00482$	&	$~~0.00196$	&	$~~0.00106$	\\
	&	$8$	&	$1$	&	$~~0.02364$	&	$~~0.00478$	&	$~~0.00195$	&	$~~0.00105$	\\
	&	$2$	&	$2$	&	$~~0.05957$	&	$~~0.01127$	&	$~~0.00454$	&	$~~0.00244$	\\
	&	$3$	&	$2$	&	$~~0.05884$	&	$~~0.01124$	&	$~~0.00454$	&	$~~0.00244$	\\
	&	$4$	&	$2$	&	$~~0.05808$	&	$~~0.01121$	&	$~~0.00453$	&	$~~0.00244$	\\
	&	$5$	&	$2$	&	$~~0.05727$	&	$~~0.01118$	&	$~~0.00453$	&	$~~0.00244$	\\
	&	$6$	&	$2$	&	$~~0.05637$	&	$~~0.01114$	&	$~~0.00452$	&	$~~0.00244$	\\
	&	$7$	&	$2$	&	$~~0.05530$	&	$~~0.01108$	&	$~~0.00452$	&	$~~0.00243$	\\
	&	$8$	&	$2$	&	$~~0.05382$	&	$~~0.01099$	&	$~~0.00450$	&	$~~0.00243$	\\
	&	$3$	&	$3$	&	$~~0.10276$	&	$~~0.01986$	&	$~~0.00803$	&	$~~0.00432$	\\
	&	$4$	&	$3$	&	$~~0.10144$	&	$~~0.01981$	&	$~~0.00802$	&	$~~0.00432$	\\
	&	$5$	&	$3$	&	$~~0.10003$	&	$~~0.01975$	&	$~~0.00801$	&	$~~0.00432$	\\
	&	$6$	&	$3$	&	$~~0.09846$	&	$~~0.01968$	&	$~~0.00800$	&	$~~0.00431$	\\
	&	$7$	&	$3$	&	$~~0.09660$	&	$~~0.01958$	&	$~~0.00799$	&	$~~0.00431$	\\
	&	$8$	&	$3$	&	$~~0.09403$	&	$~~0.01941$	&	$~~0.00796$	&	$~~0.00430$	\\
	&	$4$	&	$4$	&	$~~0.16256$	&	$~~0.03213$	&	$~~0.01304$	&	$~~0.00702$	\\
	&	$5$	&	$4$	&	$~~0.16032$	&	$~~0.03203$	&	$~~0.01302$	&	$~~0.00702$	\\
	&	$6$	&	$4$	&	$~~0.15783$	&	$~~0.03191$	&	$~~0.01301$	&	$~~0.00701$	\\
	&	$7$	&	$4$	&	$~~0.15487$	&	$~~0.03175$	&	$~~0.01299$	&	$~~0.00701$	\\
	&	$8$	&	$4$	&	$~~0.15078$	&	$~~0.03149$	&	$~~0.01294$	&	$~~0.00700$	\\
	&	$5$	&	$5$	&	$~~0.25257$	&	$~~0.05113$	&	$~~0.02084$	&	$~~0.01124$	\\
	&	$6$	&	$5$	&	$~~0.24869$	&	$~~0.05094$	&	$~~0.02081$	&	$~~0.01123$	\\
	&	$7$	&	$5$	&	$~~0.24409$	&	$~~0.05069$	&	$~~0.02078$	&	$~~0.01122$	\\
	&	$8$	&	$5$	&	$~~0.23770$	&	$~~0.05027$	&	$~~0.02071$	&	$~~0.01121$	\\
	&	$6$	&	$6$	&	$~~0.40678$	&	$~~0.08457$	&	$~~0.03466$	&	$~~0.01872$	\\
	&	$7$	&	$6$	&	$~~0.39940$	&	$~~0.08416$	&	$~~0.03461$	&	$~~0.01871$	\\
	&	$8$	&	$6$	&	$~~0.38912$	&	$~~0.08347$	&	$~~0.03450$	&	$~~0.01869$	\\
	&	$7$	&	$7$	&	$~~0.74040$	&	$~~0.15888$	&	$~~0.06563$	&	$~~0.03553$	\\
	&	$8$	&	$7$	&	$~~0.72200$	&	$~~0.15761$	&	$~~0.06543$	&	$~~0.03549$	\\
	&	$8$	&	$8$	&	$~~2.00797$	&	$~~0.45030$	&	$~~0.18848$	&	$~~0.10252$	\\
$9$	&	$	1	$	&	$	1	$	&	$	~~	0.02102	$	&	$	~~	0.00388	$	&	$	~~	0.00156	$	&	$	~~	0.00084	$	\\
	&	$	2	$	&	$	1	$	&	$	~~	0.02080	$	&	$	~~	0.00387	$	&	$	~~	0.00156	$	&	$	~~	0.00084	$	\\
	&	$	3	$	&	$	1	$	&	$	~~	0.02057	$	&	$	~~	0.00387	$	&	$	~~	0.00156	$	&	$	~~	0.00084	$	\\
	&	$	4	$	&	$	1	$	&	$	~~	0.02034	$	&	$	~~	0.00386	$	&	$	~~	0.00155	$	&	$	~~	0.00084	$	\\
	&	$	5	$	&	$	1	$	&	$	~~	0.02009	$	&	$	~~	0.00385	$	&	$	~~	0.00155	$	&	$	~~	0.00084	$	\\
	&	$	6	$	&	$	1	$	&	$	~~	0.01982	$	&	$	~~	0.00384	$	&	$	~~	0.00155	$	&	$	~~	0.00084	$	\\
	&	$	7	$	&	$	1	$	&	$	~~	0.01952	$	&	$	~~	0.00382	$	&	$	~~	0.00155	$	&	$	~~	0.00083	$	\\
	&	$	8	$	&	$	1	$	&	$	~~	0.01917	$	&	$	~~	0.00380	$	&	$	~~	0.00155	$	&	$	~~	0.00083	$	\\
	&	$	9	$	&	$	1	$	&	$	~~	0.01867	$	&	$	~~	0.00377	$	&	$	~~	0.00154	$	&	$	~~	0.00083	$	\\
	&	$	2	$	&	$	2	$	&	$	~~	0.04658	$	&	$	~~	0.00876	$	&	$	~~	0.00353	$	&	$	~~	0.00189	$	\\
	&	$	3	$	&	$	2	$	&	$	~~	0.04607	$	&	$	~~	0.00874	$	&	$	~~	0.00352	$	&	$	~~	0.00189	$	\\
	&	$	4	$	&	$	2	$	&	$	~~	0.04554	$	&	$	~~	0.00872	$	&	$	~~	0.00352	$	&	$	~~	0.00189	$	\\
	&	$	5	$	&	$	2	$	&	$	~~	0.04499	$	&	$	~~	0.00870	$	&	$	~~	0.00352	$	&	$	~~	0.00189	$	\\
	&	$	6	$	&	$	2	$	&	$	~~	0.04439	$	&	$	~~	0.00867	$	&	$	~~	0.00351	$	&	$	~~	0.00189	$	\\
	&	$	7	$	&	$	2	$	&	$	~~	0.04372	$	&	$	~~	0.00864	$	&	$	~~	0.00351	$	&	$	~~	0.00189	$	\\
	&	$	8	$	&	$	2	$	&	$	~~	0.04293	$	&	$	~~	0.00860	$	&	$	~~	0.00350	$	&	$	~~	0.00189	$	\\
	&	$	9	$	&	$	2	$	&	$	~~	0.04182	$	&	$	~~	0.00853	$	&	$	~~	0.00349	$	&	$	~~	0.00189	$	\\
	&	$	3	$	&	$	3	$	&	$	~~	0.07871	$	&	$	~~	0.01509	$	&	$	~~	0.00609	$	&	$	~~	0.00328	$	\\
	&	$	4	$	&	$	3	$	&	$	~~	0.07781	$	&	$	~~	0.01505	$	&	$	~~	0.00609	$	&	$	~~	0.00327	$	\\
	&	$	5	$	&	$	3	$	&	$	~~	0.07687	$	&	$	~~	0.01502	$	&	$	~~	0.00608	$	&	$	~~	0.00327	$	\\
	&	$	6	$	&	$	3	$	&	$	~~	0.07585	$	&	$	~~	0.01497	$	&	$	~~	0.00607	$	&	$	~~	0.00327	$	\\
	&	$	7	$	&	$	3	$	&	$	~~	0.07472	$	&	$	~~	0.01492	$	&	$	~~	0.00607	$	&	$	~~	0.00327	$	\\
	&	$	8	$	&	$	3	$	&	$	~~	0.07337	$	&	$	~~	0.01484	$	&	$	~~	0.00606	$	&	$	~~	0.00327	$	\\
	&	$	9	$	&	$	3	$	&	$	~~	0.07148	$	&	$	~~	0.01472	$	&	$	~~	0.00604	$	&	$	~~	0.00326	$	\\
	&	$	4	$	&	$	4	$	&	$	~~	0.12090	$	&	$	~~	0.02364	$	&	$	~~	0.00957	$	&	$	~~	0.00515	$	\\
	&	$	5	$	&	$	4	$	&	$	~~	0.11944	$	&	$	~~	0.02358	$	&	$	~~	0.00956	$	&	$	~~	0.00515	$	\\
	&	$	6	$	&	$	4	$	&	$	~~	0.11788	$	&	$	~~	0.02351	$	&	$	~~	0.00956	$	&	$	~~	0.00515	$	\\
	&	$	7	$	&	$	4	$	&	$	~~	0.11612	$	&	$	~~	0.02342	$	&	$	~~	0.00954	$	&	$	~~	0.00515	$	\\
	&	$	8	$	&	$	4	$	&	$	~~	0.11403	$	&	$	~~	0.02331	$	&	$	~~	0.00953	$	&	$	~~	0.00514	$	\\
	&	$	9	$	&	$	4	$	&	$	~~	0.11112	$	&	$	~~	0.02312	$	&	$	~~	0.00950	$	&	$	~~	0.00514	$	\\
	&	$	5	$	&	$	5	$	&	$	~~	0.17962	$	&	$	~~	0.03586	$	&	$	~~	0.01457	$	&	$	~~	0.00785	$	\\
	&	$	6	$	&	$	5	$	&	$	~~	0.17728	$	&	$	~~	0.03575	$	&	$	~~	0.01456	$	&	$	~~	0.00785	$	\\
	&	$	7	$	&	$	5	$	&	$	~~	0.17467	$	&	$	~~	0.03562	$	&	$	~~	0.01454	$	&	$	~~	0.00785	$	\\
	&	$	8	$	&	$	5	$	&	$	~~	0.17155	$	&	$	~~	0.03545	$	&	$	~~	0.01452	$	&	$	~~	0.00784	$	\\
	&	$	9	$	&	$	5	$	&	$	~~	0.16719	$	&	$	~~	0.03516	$	&	$	~~	0.01447	$	&	$	~~	0.00783	$	\\
	&	$	6	$	&	$	6	$	&	$	~~	0.26838	$	&	$	~~	0.05479	$	&	$	~~	0.02237	$	&	$	~~	0.01207	$	\\
	&	$	7	$	&	$	6	$	&	$	~~	0.26448	$	&	$	~~	0.05459	$	&	$	~~	0.02234	$	&	$	~~	0.01206	$	\\
	&	$	8	$	&	$	6	$	&	$	~~	0.25981	$	&	$	~~	0.05433	$	&	$	~~	0.02230	$	&	$	~~	0.01205	$	\\
	&	$	9	$	&	$	6	$	&	$	~~	0.25327	$	&	$	~~	0.05389	$	&	$	~~	0.02224	$	&	$	~~	0.01204	$	\\
	&	$	7	$	&	$	7	$	&	$	~~	0.42104	$	&	$	~~	0.08814	$	&	$	~~	0.03618	$	&	$	~~	0.01955	$	\\
	&	$	8	$	&	$	7	$	&	$	~~	0.41375	$	&	$	~~	0.08771	$	&	$	~~	0.03612	$	&	$	~~	0.01954	$	\\
	&	$	9	$	&	$	7	$	&	$	~~	0.40352	$	&	$	~~	0.08701	$	&	$	~~	0.03601	$	&	$	~~	0.01951	$	\\
	&	$	8	$	&	$	8	$	&	$	~~	0.75236	$	&	$	~~	0.16228	$	&	$	~~	0.06712	$	&	$	~~	0.03635	$	\\
	&	$	9	$	&	$	8	$	&	$	~~	0.73439	$	&	$	~~	0.16101	$	&	$	~~	0.06691	$	&	$	~~	0.03631	$	\\
	&	$	9	$	&	$	9	$	&	$	~~	2.01480	$	&	$	~~	0.45323	$	&	$	~~	0.18988	$	&	$	~~	0.10332	$	\\
$	10	$	&	$	1	$	&	$	1	$	&	$	~~	0.01709	$	&	$	~~	0.00315	$	&	$	~~	0.00126	$	&	$	~~	0.00068	$	\\
	&	$	2	$	&	$	1	$	&	$	~~	0.01693	$	&	$	~~	0.00314	$	&	$	~~	0.00126	$	&	$	~~	0.00068	$	\\
	&	$	3	$	&	$	1	$	&	$	~~	0.01676	$	&	$	~~	0.00313	$	&	$	~~	0.00126	$	&	$	~~	0.00068	$	\\
	&	$	4	$	&	$	1	$	&	$	~~	0.01659	$	&	$	~~	0.00313	$	&	$	~~	0.00126	$	&	$	~~	0.00068	$	\\
	&	$	5	$	&	$	1	$	&	$	~~	0.01641	$	&	$	~~	0.00312	$	&	$	~~	0.00126	$	&	$	~~	0.00068	$	\\
	&	$	6	$	&	$	1	$	&	$	~~	0.01622	$	&	$	~~	0.00311	$	&	$	~~	0.00126	$	&	$	~~	0.00068	$	\\
	&	$	7	$	&	$	1	$	&	$	~~	0.01601	$	&	$	~~	0.00311	$	&	$	~~	0.00126	$	&	$	~~	0.00068	$	\\
	&	$	8	$	&	$	1	$	&	$	~~	0.01578	$	&	$	~~	0.00309	$	&	$	~~	0.00126	$	&	$	~~	0.00068	$	\\
	&	$	9	$	&	$	1	$	&	$	~~	0.01551	$	&	$	~~	0.00308	$	&	$	~~	0.00125	$	&	$	~~	0.00068	$	\\
	&	$	10	$	&	$	1	$	&	$	~~	0.01512	$	&	$	~~	0.00305	$	&	$	~~	0.00125	$	&	$	~~	0.00067	$	\\
	&	$	2	$	&	$	2	$	&	$	~~	0.03743	$	&	$	~~	0.00700	$	&	$	~~	0.00282	$	&	$	~~	0.00151	$	\\
	&	$	3	$	&	$	2	$	&	$	~~	0.03706	$	&	$	~~	0.00699	$	&	$	~~	0.00282	$	&	$	~~	0.00151	$	\\
	&	$	4	$	&	$	2	$	&	$	~~	0.03668	$	&	$	~~	0.00698	$	&	$	~~	0.00281	$	&	$	~~	0.00151	$	\\
	&	$	5	$	&	$	2	$	&	$	~~	0.03628	$	&	$	~~	0.00696	$	&	$	~~	0.00281	$	&	$	~~	0.00151	$	\\
	&	$	6	$	&	$	2	$	&	$	~~	0.03586	$	&	$	~~	0.00695	$	&	$	~~	0.00281	$	&	$	~~	0.00151	$	\\
	&	$	7	$	&	$	2	$	&	$	~~	0.03541	$	&	$	~~	0.00693	$	&	$	~~	0.00281	$	&	$	~~	0.00151	$	\\
	&	$	8	$	&	$	2	$	&	$	~~	0.03490	$	&	$	~~	0.00690	$	&	$	~~	0.00280	$	&	$	~~	0.00151	$	\\
	&	$	9	$	&	$	2	$	&	$	~~	0.03429	$	&	$	~~	0.00687	$	&	$	~~	0.00280	$	&	$	~~	0.00151	$	\\
	&	$	10	$	&	$	2	$	&	$	~~	0.03344	$	&	$	~~	0.00681	$	&	$	~~	0.00279	$	&	$	~~	0.00151	$	\\
	&	$	3	$	&	$	3	$	&	$	~~	0.06229	$	&	$	~~	0.01186	$	&	$	~~	0.00478	$	&	$	~~	0.00257	$	\\
	&	$	4	$	&	$	3	$	&	$	~~	0.06165	$	&	$	~~	0.01184	$	&	$	~~	0.00478	$	&	$	~~	0.00257	$	\\
	&	$	5	$	&	$	3	$	&	$	~~	0.06098	$	&	$	~~	0.01181	$	&	$	~~	0.00478	$	&	$	~~	0.00257	$	\\
	&	$	6	$	&	$	3	$	&	$	~~	0.06028	$	&	$	~~	0.01178	$	&	$	~~	0.00477	$	&	$	~~	0.00257	$	\\
	&	$	7	$	&	$	3	$	&	$	~~	0.05952	$	&	$	~~	0.01175	$	&	$	~~	0.00477	$	&	$	~~	0.00257	$	\\
	&	$	8	$	&	$	3	$	&	$	~~	0.05867	$	&	$	~~	0.01171	$	&	$	~~	0.00476	$	&	$	~~	0.00257	$	\\
	&	$	9	$	&	$	3	$	&	$	~~	0.05765	$	&	$	~~	0.01165	$	&	$	~~	0.00475	$	&	$	~~	0.00256	$	\\
	&	$	10	$	&	$	3	$	&	$	~~	0.05621	$	&	$	~~	0.01156	$	&	$	~~	0.00474	$	&	$	~~	0.00256	$	\\
	&	$	4	$	&	$	4	$	&	$	~~	0.09369	$	&	$	~~	0.01816	$	&	$	~~	0.00734	$	&	$	~~	0.00395	$	\\
	&	$	5	$	&	$	4	$	&	$	~~	0.09269	$	&	$	~~	0.01812	$	&	$	~~	0.00734	$	&	$	~~	0.00395	$	\\
	&	$	6	$	&	$	4	$	&	$	~~	0.09162	$	&	$	~~	0.01808	$	&	$	~~	0.00733	$	&	$	~~	0.00395	$	\\
	&	$	7	$	&	$	4	$	&	$	~~	0.09047	$	&	$	~~	0.01802	$	&	$	~~	0.00733	$	&	$	~~	0.00395	$	\\
	&	$	8	$	&	$	4	$	&	$	~~	0.08918	$	&	$	~~	0.01796	$	&	$	~~	0.00732	$	&	$	~~	0.00395	$	\\
	&	$	9	$	&	$	4	$	&	$	~~	0.08764	$	&	$	~~	0.01787	$	&	$	~~	0.00730	$	&	$	~~	0.00394	$	\\
	&	$	10	$	&	$	4	$	&	$	~~	0.08547	$	&	$	~~	0.01773	$	&	$	~~	0.00728	$	&	$	~~	0.00394	$	\\
	&	$	5	$	&	$	5	$	&	$	~~	0.13511	$	&	$	~~	0.02668	$	&	$	~~	0.01082	$	&	$	~~	0.00583	$	\\
	&	$	6	$	&	$	5	$	&	$	~~	0.13357	$	&	$	~~	0.02661	$	&	$	~~	0.01081	$	&	$	~~	0.00583	$	\\
	&	$	7	$	&	$	5	$	&	$	~~	0.13191	$	&	$	~~	0.02653	$	&	$	~~	0.01080	$	&	$	~~	0.00582	$	\\
	&	$	8	$	&	$	5	$	&	$	~~	0.13004	$	&	$	~~	0.02644	$	&	$	~~	0.01079	$	&	$	~~	0.00582	$	\\
	&	$	9	$	&	$	5	$	&	$	~~	0.12780	$	&	$	~~	0.02631	$	&	$	~~	0.01077	$	&	$	~~	0.00582	$	\\
	&	$	10	$	&	$	5	$	&	$	~~	0.12464	$	&	$	~~	0.02610	$	&	$	~~	0.01074	$	&	$	~~	0.00581	$	\\
	&	$	6	$	&	$	6	$	&	$	~~	0.19298	$	&	$	~~	0.03885	$	&	$	~~	0.01582	$	&	$	~~	0.00853	$	\\
	&	$	7	$	&	$	6	$	&	$	~~	0.19060	$	&	$	~~	0.03874	$	&	$	~~	0.01580	$	&	$	~~	0.00852	$	\\
	&	$	8	$	&	$	6	$	&	$	~~	0.18792	$	&	$	~~	0.03861	$	&	$	~~	0.01578	$	&	$	~~	0.00852	$	\\
	&	$	9	$	&	$	6	$	&	$	~~	0.18470	$	&	$	~~	0.03842	$	&	$	~~	0.01576	$	&	$	~~	0.00851	$	\\
	&	$	10	$	&	$	6	$	&	$	~~	0.18017	$	&	$	~~	0.03812	$	&	$	~~	0.01571	$	&	$	~~	0.00850	$	\\
	&	$	7	$	&	$	7	$	&	$	~~	0.28076	$	&	$	~~	0.05773	$	&	$	~~	0.02360	$	&	$	~~	0.01274	$	\\
	&	$	8	$	&	$	7	$	&	$	~~	0.27686	$	&	$	~~	0.05753	$	&	$	~~	0.02358	$	&	$	~~	0.01273	$	\\
	&	$	9	$	&	$	7	$	&	$	~~	0.27218	$	&	$	~~	0.05726	$	&	$	~~	0.02354	$	&	$	~~	0.01272	$	\\
	&	$	10	$	&	$	7	$	&	$	~~	0.26557	$	&	$	~~	0.05680	$	&	$	~~	0.02347	$	&	$	~~	0.01271	$	\\
	&	$	8	$	&	$	8	$	&	$	~~	0.43216	$	&	$	~~	0.09100	$	&	$	~~	0.03740	$	&	$	~~	0.02022	$	\\
	&	$	9	$	&	$	8	$	&	$	~~	0.42499	$	&	$	~~	0.09057	$	&	$	~~	0.03734	$	&	$	~~	0.02021	$	\\
	&	$	10	$	&	$	8	$	&	$	~~	0.41485	$	&	$	~~	0.08986	$	&	$	~~	0.03723	$	&	$	~~	0.02018	$	\\
	&	$	9	$	&	$	9	$	&	$	~~	0.76159	$	&	$	~~	0.16499	$	&	$	~~	0.06832	$	&	$	~~	0.03702	$	\\
	&	$	10	$	&	$	9	$	&	$	~~	0.74402	$	&	$	~~	0.16373	$	&	$	~~	0.06811	$	&	$	~~	0.03697	$	\\
	&	$	10	$	&	$	10	$	&	$	~~	2.01965	$	&	$	~~	0.45554	$	&	$	~~	0.19101	$	&	$	~~	0.10397	$	\\
\hline
\end{longtable}
\vspace{-0.1cm}
Here, the condition $\sum_{r=1}^n\sum_{s=1}^n\sigma_{r,s:n}=n\sigma^2$ (David and Nagaraja, \cite{david}), is satisfied, where $\sigma_{r,s:n}=\text{Cov}(X_{r:n},X_{s:n})$ and $\sigma^2=\text{Var}(X)$.

\section{Linear Estimation}
Suppose $X_{1:n} \leq X_{2:n} \leq ... \leq X_{n-m:n},~m=0,1,..., n-1,$ denote  Type-II  censored order statistics  from $XL(\varphi,\sigma,\psi)$.
 Besides, let $\text{Z}_{r:n} =\dfrac{X_{r:n}-\varphi}{\sigma}$, $\text{E}(\text{Z}_{r:n})=\varphi^{(1)}_{r:n},~1 \leq  r \leq n-m,$ and $\text{Cov}(\text{Z}_{r:n}, \text{Z}_{s:n})=\sigma_{r,s:n}=\varphi^{(1,1)}_{r,s:n}-\varphi^{(1)}_{r:n}\varphi^{(1)}_{s:n}, 1 \leq r < s \leq n-m$.\\

We shall use the following notations:
\begin{equation*}
{\bf X}=(X_{1:n}, X_{2:n}, ..., X_{n-m:n})^{\text{T}},
\end{equation*}
\vspace{-0.1cm}
\begin{equation*}
{\boldsymbol \varphi} =(\varphi_{1:n}, \varphi_{2:n}, ..., \varphi_{n-m:n})^{\text{T}},
\end{equation*}
\vspace{-0.1cm}
\begin{equation*}
\textbf{1}=(1,1,...,1)^{\text{T}}_{1\times (n-m)},
\end{equation*}
and
\begin{equation*}
{\bf{\Sigma}}= \left( \left( \sigma_{r,s:n}\right) \right),~ 1 \leq r,~s \leq n-m.
\end{equation*}

\vspace{0.3cm}
Then, the BLUEs of $\varphi$ and $\sigma$, denoted by $\widehat{\varphi}$ and $\widehat{\sigma}$, respectively, are given by *
\begin{align}\label{aaa}
\widehat{\varphi}= \left\lbrace \frac{{\boldsymbol \varphi}^{\text{T}}{\bf{\Sigma}}^{-1}{\boldsymbol \varphi}  \textbf{1}^{\text{T}}{\bf{\Sigma}}^{-1}-{\boldsymbol \varphi}^{\text{T}} {\bf{\Sigma}}^{-1}\textbf{1}{\boldsymbol \varphi}^{\text{T}}{\bf{\Sigma}}^{-1}}{({\boldsymbol \varphi}^{\text{T}}{\bf{\Sigma}}^{-1}{\boldsymbol \varphi})(\textbf{1}^{\text{T}}{\bf{\Sigma}}^{-1}\textbf{1})-({\boldsymbol \varphi}^{\text{T}}{\bf{\Sigma}}^{-1}\textbf{1})^{2}}\right\rbrace {\bf X}~ = \sum_{r=1}^{n-m}a_r X_{r:n},
\end{align}
and
\begin{align}\label{bbb}
\widehat{\sigma}= \left\lbrace \frac{\textbf{1}^{\text{T}}{\bf{\Sigma}}^{-1}\textbf{1} {\boldsymbol \varphi}^{\text{T}}{\bf{\Sigma}}^{-1}-\textbf{1}^{\text{T}} {\bf{\Sigma}}^{-1}{\boldsymbol \varphi}\textbf{1}^{\text{T}}{\bf{\Sigma}}^{-1}}{({\boldsymbol \varphi}^{\text{T}}{\bf{\Sigma}}^{-1}{\boldsymbol \varphi})(\textbf{1}^{\text{T}}{\bf{\Sigma}}^{-1}\textbf{1})-({\boldsymbol \varphi}^{\text{T}}{\bf{\Sigma}}^{-1}\textbf{1})^{2}}\right\rbrace {\bf X}~ = \sum_{r=1}^{n-m}b_r X_{r:n}.
\end{align}

\vspace{0.4cm}
The variances and covariance of these BLUEs are given by

\begin{align*}
\text{Var}(\widehat{\varphi})= \sigma^{2}\left\lbrace \frac{{\boldsymbol \varphi}^{\text{T}}{\bf{\Sigma}}^{-1}{\boldsymbol \varphi}}{({\boldsymbol \varphi}^{\text{T}}{\bf{\Sigma}}^{-1}{\boldsymbol \varphi})(\textbf{1}^{\text{T}}{\bf{\Sigma}}^{-1}\textbf{1})-({\boldsymbol \varphi}^{\text{T}}{\bf{\Sigma}}^{-1}\textbf{1})^{2}}\right\rbrace = \sigma^{2} V_1,
\end{align*}

\begin{align*}
\text{Var}(\widehat{\sigma})= \sigma^{2}\left\lbrace \frac{\textbf{1}^{\text{T}}{\bf{\Sigma}}^{-1}\textbf{1}}{({\boldsymbol \varphi} ^{\text{T}}{\bf{\Sigma}}^{-1}{\boldsymbol \varphi} )(\textbf{1}^{\text{T}}{\bf{\Sigma}}^{-1}\textbf{1})-({\boldsymbol \varphi} ^{\text{T}}{\bf{\Sigma}}^{-1}\textbf{1})^{2}}\right\rbrace = \sigma^{2} V_2
\end{align*}

\vspace{0.3cm}
and
\begin{align*}
\text{Cov}(\widehat{\varphi},~\widehat{\sigma})= \sigma^{2}\left\lbrace \frac{-{\boldsymbol \varphi}^{\text{T}}{\bf{\Sigma}}^{-1}\textbf{1}}{({\boldsymbol \varphi}^{\text{T}}{\bf{\Sigma}}^{-1}{\boldsymbol \varphi})(\textbf{1}^{\text{T}}{\bf{\Sigma}}^{-1}\textbf{1})-({\boldsymbol \varphi}^{\text{T}}{\bf{\Sigma}}^{-1}\textbf{1})^{2}}\right\rbrace = \sigma^{2} V_3,
\end{align*}

\vspace{0.3cm}
respectively. See for details, Lloyd \cite{Lloyd}, Arnold et al. \cite{arnd}, Balakrishnan and Cohen \cite{BR} and David and Nagaraja \cite{david}.\\

We proceed to derive the BLIEs of $\varphi$ and $\sigma$. In view of Mann \cite{Man},  the BLIEs of $\varphi$ and $\sigma$ are given by
\begin{align}\label{6.6}
\widetilde{\varphi}= \left\lbrace \frac{\left({\boldsymbol \varphi}^{\text{T}}{\bf{\Sigma}}^{-1}{\boldsymbol \varphi}+1\right)  \textbf{1}^{\text{T}}{\bf{\Sigma}}^{-1}-{\boldsymbol \varphi}^{\text{T}} {\bf{\Sigma}}^{-1}\textbf{1}{\boldsymbol \varphi}^{\text{T}}{\bf{\Sigma}}^{-1}}{({\boldsymbol \varphi}^{\text{T}}{\bf{\Sigma}}^{-1}{\boldsymbol \varphi})(\textbf{1}^{\text{T}}{\bf{\Sigma}}^{-1}\textbf{1})-({\boldsymbol \varphi}^{\text{T}}{\bf{\Sigma}}^{-1}\textbf{1})^{2}+\textbf{1}^{\text{T}}{\bf{\Sigma}}^{-1}\textbf{1}}\right\rbrace {\bf X}~ = \widehat{\varphi}-\dfrac{V_3}{1+V_2}\widehat{\sigma}= \sum_{r=1}^{n-m}a'_r X_{r:n},
\end{align}
\begin{align}\label{6.7}
\widetilde{\sigma}= \left\lbrace \frac{\textbf{1}^{\text{T}}{\bf{\Sigma}}^{-1}\textbf{1} {\boldsymbol \varphi}^{\text{T}}{\bf{\Sigma}}^{-1}-\textbf{1}^{\text{T}} {\bf{\Sigma}}^{-1}{\boldsymbol \varphi}\textbf{1}^{\text{T}}{\bf{\Sigma}}^{-1}}{({\boldsymbol \varphi}^{\text{T}}{\bf{\Sigma}}^{-1}{\boldsymbol \varphi})(\textbf{1}^{\text{T}}{\bf{\Sigma}}^{-1}\textbf{1})-({\boldsymbol \varphi}^{\text{T}}{\bf{\Sigma}}^{-1}\textbf{1})^{2}+\textbf{1}^{\text{T}}{\bf{\Sigma}}^{-1}\textbf{1}}\right\rbrace {\bf X}~ = \dfrac{\widehat{\sigma}}{1+V_2}= \sum_{r=1}^{n-m}b'_r X_{r:n},
\end{align}
where $V_1=\frac{1}{\sigma^2}\text{Var}(\widehat{\varphi}),$
$V_2=\frac{1}{\sigma^2}\text{Var}(\widehat{\sigma})$ and
$V_3=\frac{1}{\sigma^2}\text{Cov}(\widehat{\varphi},
\widehat{\sigma}).$

Next, using (\ref{aaa}) and (\ref{bbb}), respectively, we calculated the coefficients of the BLUEs for location and scale parameters for the Type-II right censored samples, presented in Tables 3 and 4 for   $n=6$ and 10 with $m$ ranging from 0 to  $[n/2]-1$ and $\psi=1,\cdots,4$. Moreover, Tables 5 and 6 present the coefficients of the BLIEs using equations (\ref{6.6}) and (\ref{6.7}), respectively.

\vspace{0.3cm}
Note that the following conditions verify the validity of the computed coefficients:
$$\sum_{r=1}^{n-m}a_r=1\quad \text{and}\quad \sum_{r=1}^{n-m}b_r=0.$$

Based on the BLUEs for the location and scale parameters, confidence intervals (CIs) for $\varphi$ and $\sigma$ can be derived using the following pivotal quantities:
\[
T_1 = \frac{\hat\varphi - \varphi}{\hat\sigma \sqrt{V_1}},
\quad \text{and} \quad
T_2 = \frac{\hat\sigma - \sigma}{\sigma \sqrt{V_2}}.
\]

To construct these CIs, we need the percentage points of $T_1$ and $T_2$, which can be calculated using the BLUEs $\varphi^*$ and $\sigma^*$ through the Monte Carlo method. Table 7 presents these percentage points, determined from 10,000 simulations across various values of $n$ and $\psi$. Using these simulated percentage points, we can construct a $100(1 - \Upsilon)\%$ CI for $\varphi$ based on the pivotal quantity $T_1$ as follows:
\[
\text{P}\left( \hat\varphi - \hat\sigma \sqrt{V_1} T_1(1 - \Upsilon/2) \leq \varphi \leq \hat\varphi - \hat\sigma \sqrt{V_1} T_1(\Upsilon/2) \right) = 1 -\Upsilon,
\]
where $T_1(\tau)$ denotes the left percentage point of $T_1$ at $\tau$, i.e. $\text{P}(T_1 < T_1(\tau)) = \tau$.

Similarly, a $100(1 - \Upsilon)\%$ CI for $\sigma$ can be constructed using the pivotal quantity $T_2$ as follows:
\[
\text{P}\left( \frac{\hat\sigma}{1 + \sqrt{V_2} T_2(1 - \Upsilon/2)} \leq \sigma \leq \frac{\hat\sigma}{1 + \sqrt{V_2} T_2(\Upsilon/2)} \right) = 1 - \Upsilon,
\]
where $T_2(\tau)$ is the left percentage point of $T_2$ at $\tau$, i.e. $\text{P}(T_2 < T_2(\tau)) = \tau$.

\vspace{0.2cm}
Based on the BLIEs, we can construct confidence intervals (CIs) for the location and scale parameters using the following pivotal quantities:
\[
T_3 = \frac{\widetilde{\varphi} - \varphi}{\widetilde{\sigma} \sqrt{V_1 - \frac{V_3^2 (2 + V_2)}{(1 + V_2)^2}}},\quad 
\quad \text{and} \quad
T_4 = \frac{\widetilde{\sigma} - \sigma}{\sigma \frac{\sqrt{V_2}}{1 + V_2}}.
\]

Table \ref{tabcri2} shows the percentage points of \(T_3\) and \(T_4\) based on 10,000 simulations with various choices of \(n\) and \(\psi\). Using the BLIEs and the values from Table \ref{tabcri2}, we can construct a \(100(1 - \Upsilon)\%\) CI for \(\varphi\) through the pivotal quantity \(T_3\) as follows:
\[
\text{P}\left( \widetilde{\varphi} - \widetilde{\sigma} \sqrt{V_1 - \frac{V_3^2 (2 + V_2)}{(1 + V_2)^2}} T_3(1 - \Upsilon/2) \leq \varphi \leq \widetilde{\varphi} - \widetilde{\sigma} \sqrt{V_1 - \frac{V_3^2 (2 + V_2)}{(1 + V_2)^2}} T_3(\Upsilon/2) \right) = 1 - \Upsilon,
\]
where \(T_3(\tau)\) represents the left percentage point of \(T_3\) at \(\tau\), i.e., \(\text{P}(T_3 < T_3(\tau)) = \tau\).

\vspace{0.20cm}
Similarly, a \(100(1 - \Upsilon)\%\) CI for \(\sigma\) can be constructed using the pivotal quantity \(T_4\) as follows:
\[
\text{P}\left( \frac{\widetilde{\sigma}}{1 + \frac{\sqrt{V_2}}{1 + V_2} T_4(1 - \Upsilon/2)} \leq \sigma \leq \frac{\widetilde{\sigma}}{1 + \frac{\sqrt{V_2}}{1 + V_2} T_4(\Upsilon/2)} \right) = 1 - \Upsilon,
\]
where \(T_4(\tau)\) denotes the left percentage point of \(T_4\) at \(\tau\), i.e., \(\text{P}(T_4 < T_4(\tau)) = \tau\).

\newpage
\small{\bf Table 3:} Coefficients for the BLUEs of the location parameter.
\vspace{-0.3cm}
\setlength{\tabcolsep}{1.33em}
\small
\begin{longtable}{|r|c|c|r|r|r|r|r|}
\multicolumn{8}{r}{{({\em Continued})}}\\[.9ex]
\endfoot
\endlastfoot
\hline
\multicolumn{1}{|l|}{$~\psi$} & $n$     & $m$     & \multicolumn{5}{c|}{$a_i$,~$i =1,2,3,\ldots,(n-m)$} \\
\hline
$1.0$	&	$6$	&	$0$	&	$1.16480$	&	$-0.02847$	&	$-0.03056$	&	$-0.03268$	&	$-0.03497$	\\
		&		&		&	$-0.03813$	&				&				&				&				\\
		&		&	$1$	&	$1.20851$	&	$-0.03658$	&	$-0.03907$	&	$-0.04158$	&	$-0.09128$	\\
		&		&	$2$	&	$1.28036$	&	$-0.05002$	&	$-0.05317$	&	$-0.17717$	&				\\
		&  $10$	&	$0$	&	$1.09713$	&	$-0.00887$	&	$-0.00933$	&	$-0.00980$	&	$-0.01025$	\\
		&		&		&	$-0.01072$	&	$-0.01119$	&	$-0.01168$	&	$-0.01224$	&	$-0.01306$	\\
		&		&	$1$	&	$1.11032$	&	$-0.01019$	&	$-0.01070$	&	$-0.01120$	&	$-0.01170$	\\
		&		&		&	$-0.01221$	&	$-0.01272$	&	$-0.01326$	&	$-0.02834$	&				\\
		&		&	$2$	&	$1.12709$	&	$-0.01188$	&	$-0.01245$	&	$-0.01301$	&	$-0.01356$	\\
		&		&		&	$-0.01412$	&	$-0.01469$	&	$-0.04738$	&				&				\\
		&		&	$3$	&	$1.14925$	&	$-0.01413$	&	$-0.01477$	&	$-0.01541$	&	$-0.01604$	\\
		&		&		&	$-0.01666$	&	$-0.07225$	&				&				&				\\
		&		&	$4$	&	$1.18006$	&	$-0.01727$	&	$-0.01802$	&	$-0.01876$	&	$-0.01948$	\\
		&		&		&	$-0.10654$	&				&				&				&				\\
$2.0$	&	$6$	&	$0$	&	$1.16595$	&	$-0.03235$	&	$-0.03265$	&	$-0.03301$	&	$-0.03348$	\\
		&		&		&	$-0.03447$	&				&				&				&				\\
		&		&	$1$	&	$1.20814$	&	$-0.04066$	&	$-0.04103$	&	$-0.04147$	&	$-0.08498$	\\
		&		&	$2$	&	$1.27808$	&	$-0.05447$	&	$-0.05494$	&	$-0.16867$	&				\\
		& $10$	&	$0$	&	$1.09930$	&	$-0.01072$	&	$-0.01078$	&	$-0.01084$	&	$-0.0109$	\\
		&		&		&	$-0.01097$	&	$-0.01106$	&	$-0.01116$	&	$-0.01129$	&	$-0.01158$	\\
		&		&	$1$	&	$1.11200$	&	$-0.01211$	&	$-0.01217$	&	$-0.01224$	&	$-0.01231$	\\
		&		&		&	$-0.01239$	&	$-0.01248$	&	$-0.01259$	&	$-0.02571$	&				\\
		&		&	$2$	&	$1.12824$	&	$-0.01388$	&	$-0.01395$	&	$-0.01403$	&	$-0.01411$	\\
		&		&		&	$-0.01420$	&	$-0.01430$	&	$-0.04376$	&				&				\\
		&		&	$3$	&	$1.14982$	&	$-0.01625$	&	$-0.01633$	&	$-0.01641$	&	$-0.0165$	\\
		&		&		&	$-0.01661$	&	$-0.06772$	&				&				&				\\
		&		&	$4$	&	$1.17996$	&	$-0.01955$	&	$-0.01964$	&	$-0.01974$	&	$-0.01985$	\\
		&		&		&	$-0.10118$	&				&				&				&				\\
$3.0$	&	$6$	&	$0$	&	$1.16638$	&	$-0.03301$	&	$-0.03309$	&	$-0.0332$	&	$-0.03335$	\\
		&		&		&	$-0.03373$	&				&				&				&				\\
		&		&	$1$	&	$1.20824$	&	$-0.04135$	&	$-0.04145$	&	$-0.04158$	&	$-0.08387$	\\
		&		&	$2$	&	$1.27785$	&	$-0.05521$	&	$-0.05535$	&	$-0.16729$	&				\\
		& $10$	&	$0$	&	$1.09975$	&	$-0.01099$	&	$-0.01100$	&	$-0.01102$	&	$-0.01104$	\\
		&		&		&	$-0.01106$	&	$-0.01108$	&	$-0.01112$	&	$-0.01116$	&	$-0.01128$	\\
		&		&	$1$	&	$1.11233$	&	$-0.01238$	&	$-0.01240$	&	$-0.01242$	&	$-0.01244$	\\
		&		&		&	$-0.01246$	&	$-0.01249$	&	$-0.01252$	&	$-0.02524$	&				\\
		&		&	$2$	&	$1.12846$	&	$-0.01417$	&	$-0.01418$	&	$-0.01420$	&	$-0.01423$	\\
		&		&		&	$-0.01425$	&	$-0.01428$	&	$-0.04315$	&				&				\\
		&		&	$3$	&	$1.149940$	&	$-0.01654$	&	$-0.01656$	&	$-0.01658$	&	$-0.01661$	\\
		&		&		&	$-0.01664$	&	$-0.06699$	&				&				&				\\
		&		&	$4$	&	$1.17998$	&	$-0.01987$	&	$-0.01989$	&	$-0.01992$	&	$-0.01995$	\\
		&		&		&	$-0.10036$	&				&				&				&				\\
$4.0$	&	$6$	&	$0$	&	$1.16654$	&	$-0.0332$	&	$-0.03323$	&	$-0.03327$	&	$-0.03333$	\\
		&		&		&	$-0.03350$	&				&				&				&				\\
		&		&	$1$	&	$1.20829$	&	$-0.04154$	&	$-0.04158$	&	$-0.04162$	&	$-0.08356$	\\
		&		&	$2$	&	$1.27780$	&	$-0.05542$	&	$-0.05547$	&	$-0.16692$	&				\\
		& $10$	&	$0$	&	$1.09989$	&	$-0.01106$	&	$-0.01107$	&	$-0.01107$	&	$-0.01108$	\\
		&		&		&	$-0.01109$	&	$-0.01110$	&	$-0.01111$	&	$-0.01113$	&	$-0.01118$	\\
		&		&	$1$	&	$1.11243$	&	$-0.01245$	&	$-0.01246$	&	$-0.01246$	&	$-0.01247$	\\
		&		&		&	$-0.01248$	&	$-0.01249$	&	$-0.01251$	&	$-0.02510$	&				\\
		&		&	$2$	&	$1.12852$	&	$-0.01424$	&	$-0.01424$	&	$-0.01425$	&	$-0.01426$	\\
		&		&		&	$-0.01427$	&	$-0.01428$	&	$-0.04298$	&				&				\\
		&		&	$3$	&	$1.149970$	&	$-0.01662$	&	$-0.01663$	&	$-0.01663$	&	$-0.01664$	\\
		&		&		&	$-0.01665$	&	$-0.06680$	&				&				&				\\
		&		&		&	$1.17999$	&	$-0.01995$	&	$-0.01996$	&	$-0.01997$	&	$-0.01998$	\\
		&		&		&	$-0.10014$	&				&				&				&				\\	
\hline
\end{longtable}

\setlength{\tabcolsep}{1.55em}
\small{\bf Table 4:} Coefficients for the BLUEs of the scale parameter.
\vspace{-0.2cm}
\small
\begin{longtable}{|r|c|c|r|r|r|r|r|}
\multicolumn{8}{r}{{({\em Continued})}}\\[.7ex]
\endfoot
\endlastfoot
\hline
\multicolumn{1}{|l|}{$~\psi$} & $n$        & $m$          & \multicolumn{5}{c|}{$b_i$,~$i =1,2,3,\ldots,(n-m)$} \\
\hline
$1$	&	$6$	&	$0$	&	$-0.76791$	&	$0.13836$	&	$0.14548$	&	$0.15269$	&	$0.16041$	\\	
	&		&		&	$0.17098$	&				&				&				&				\\
	&		&	$1$	&	$-0.96393$	&	$0.17471$	&	$0.18364$	&	$0.19262$	&	$0.41296$	\\
	&		&	$2$	&	$-1.28900$	&	$0.23553$	&	$0.24744$	&	$0.80603$	&				\\
	& $10$	&	$0$	&	$-0.75637$	&	$0.07371$	&	$0.07623$	&	$0.07872$	&	$0.08118$	\\	
	&		&		&	$0.08366$	&	$0.08618$	&	$0.08882$	&	$0.09177$	&	$0.09611$	\\
	&		&	$1$	&	$-0.85342$	&	$0.08342$	&	$0.08627$	&	$0.08908$	&	$0.09186$	\\	
	&		&		&	$0.09465$	&	$0.09748$	&	$0.10042$	&	$0.21025$	&				\\
	&		&	$2$	&	$-0.97783$	&	$0.09597$	&	$0.09925$	&	$0.10247$	&	$0.10566$	\\
	&		&		&	$0.10884$	&	$0.11205$	&	$0.35360$	&				&				\\
	&		&	$3$	&	$-1.14321$	&	$0.11276$	&	$0.11659$	&	$0.12036$	&	$0.12409$	\\
	&		&		&	$0.12780$	&	$0.54161$	&				&				&				\\
	&		&	$4$	&	$-1.37418$	&	$0.13631$	&	$0.14093$	&	$0.14547$	&	$0.14994$	\\
	&		&		&	$0.80153$	&				&				&				&				\\
$2$	&	$6$	&	$0$	&	$-1.77763$	&	$0.34748$	&	$0.35034$	&	$0.35377$	&	$0.35831$	\\
	&		&		&	$0.36774$	&				&				&				&				\\
	&		&	$1$	&	$-2.22777$	&	$0.43621$	&	$0.43978$	&	$0.44404$	&	$0.90775$	\\		
	&		&	$2$	&	$-2.97486$	&	$0.58368$	&	$0.58841$	&	$1.80277$	&				\\	
	& $10$	&	$0$	&	$-1.77207$	&	$0.19201$	&	$0.19286$	&	$0.19379$	&	$0.19481$	\\
	&		&		&	$0.19595$	&	$0.19727$	&	$0.19886$	&	$0.20099$	&	$0.20551$	\\
	&		&	$1$	&	$-1.99741$	&	$0.21662$	&	$0.21758$	&	$0.21863$	&	$0.21978$	\\
	&		&		&	$0.22106$	&	$0.22254$	&	$0.22431$	&	$0.45688$	&				\\
	&		&	$2$	&	$-2.28585$	&	$0.24817$	&	$0.24927$	&	$0.25047$	&	$0.25178$	\\
	&		&		&	$0.25324$	&	$0.25492$	&	$0.77801$	&				&				\\
	&		&	$3$	&	$-2.66951$	&	$0.29016$	&	$0.29145$	&	$0.29284$	&	$0.29437$	\\
	&		&		&	$0.29607$	&	$1.20461$	&				&				&				\\
	&		&	$4$	&	$-3.20583$	&	$0.34890$	&	$0.35044$	&	$0.35211$	&	$0.35394$	\\
	&		&		&	$1.80043$	&				&				&				&				\\
$3$	&	$6$	&	$0$	&	$-2.81058$	&	$0.55792$	&	$0.55922$	&	$0.56088$	&	$0.56322$	\\
	&		&		&	$0.56933$	&				&				&				&				\\
	&		&	$1$	&	$-3.51722$	&	$0.69863$	&	$0.70026$	&	$0.70231$	&	$1.41602$	\\	
	&		&	$2$	&	$-4.69249$	&	$0.93273$	&	$0.93489$	&	$2.82487$	&				\\
	& $10$	&	$0$	&	$-2.80782$	&	$0.30953$	&	$0.30990$	&	$0.31032$	&	$0.31079$	\\
	&		&		&	$0.31134$	&	$0.31199$	&	$0.31282$	&	$0.31401$	&	$0.31712$	\\
	&		&	$1$	&	$-3.16153$	&	$0.34864$	&	$0.34906$	&	$0.34953$	&	$0.35006$	\\
	&		&		&	$0.35067$	&	$0.35141$	&	$0.35233$	&	$0.70983$	&				\\
	&		&	$2$	&	$-3.61518$	&	$0.39882$	&	$0.39930$	&	$0.39983$	&	$0.40044$	\\
	&		&		&	$0.40114$	&	$0.40197$	&	$1.21369$	&				&				\\
	&		&	$3$	&	$-4.21937$	&	$0.46566$	&	$0.46622$	&	$0.46684$	&	$0.46754$	\\
	&		&		&	$0.46836$	&	$1.88476$	&				&				&				\\
	&		&	$4$	&	$-5.06468$	&	$0.55918$	&	$0.55985$	&	$0.56060$	&	$0.56144$	\\
	&		&		&	$2.82361$	&				&				&				&				\\
$4$	&	$6$	&	$0$	&	$-3.83839$	&	$0.76529$	&	$0.76598$	&	$0.76688$	&	$0.76818$	\\
	&		&		&	$0.77206$	&				&				&				&				\\
	&		&	$1$	&	$-4.80059$	&	$0.95740$	&	$0.95826$	&	$0.95937$	&	$1.92556$	\\	
	&		&	$2$	&	$-6.40255$	&	$1.27726$	&	$1.27841$	&	$3.84688$	&				\\	
	& $10$	&	$0$	&	$-3.83686$	&	$0.42495$	&	$0.42514$	&	$0.42536$	&	$0.42562$	\\
	&		&		&	$0.42591$	&	$0.42627$	&	$0.42674$	&	$0.42742$	&	$0.42945$	\\
	&		&	$1$	&	$-4.31826$	&	$0.47834$	&	$0.47856$	&	$0.47880$	&	$0.47909$	\\
	&		&		&	$0.47942$	&	$0.47982$	&	$0.48034$	&	$0.96389$	&				\\
	&		&	$2$	&	$-4.93642$	&	$0.54690$	&	$0.54715$	&	$0.54743$	&	$0.54776$	\\
	&		&		&	$0.54813$	&	$0.54859$	&	$1.65045$	&				&				\\
	&		&	$3$	&	$-5.76016$	&	$0.63827$	&	$0.63856$	&	$0.63889$	&	$0.63927$	\\
	&		&		&	$0.63971$	&	$2.56546$	&				&				&				\\
	&		&	$4$	&	$-6.91305$	&	$0.76615$	&	$0.76650$	&	$0.76690$	&	$0.76735$	\\
	&		&		&	$3.84616$	&				&				&				&				\\
\hline
\end{longtable}

\small{\bf Table 5:} Coefficients for the BLIEs of the location parameter.
\vspace{-0.3cm}
\setlength{\tabcolsep}{1.37em}
\small
\begin{longtable}{|r|c|c|r|r|r|r|r|}
\multicolumn{8}{r}{{({\em Continued})}}\\[.8ex]
\endfoot
\endlastfoot
\hline
\multicolumn{1}{|l|}{$~\psi$} & $n$     & $m$     & \multicolumn{5}{c|}{$a'_i$,~$i =1,2,3,\ldots,(n-m)$} \\
\hline
$1$	&	$6$	&	$0$	&	$1.13734$	&	$-0.02352$	&	$-0.02535$	&	$-0.02722$	&	$-0.02923$	\\
	&		&		&	$-0.03201$	&				&				&				&				\\
	&		&	$1$	&	$1.16691$	&	$-0.02904$	&	$-0.03114$	&	$-0.03327$	&	$-0.07346$	\\
	&		&	$2$	&	$1.21050$	&	$-0.03726$	&	$-0.03976$	&	$-0.13349$  &               \\
	& $10$	&	$0$	&	$1.08734$	&	$-0.00791$	&	$-0.00835$	&	$-0.00878$	&	$-0.0092$	\\
	&		&		&	$-0.00963$	&	$-0.01007$	&	$-0.01053$	&	$-0.01105$	&	$-0.01182$	\\
	&		&	$1$	&	$1.09800$	&	$-0.00898$	&	$-0.00945$	&	$-0.00992$	&	$-0.01038$	\\
	&		&		&	$-0.01084$	&	$-0.01132$	&	$-0.01181$	&	$-0.0253$	&				\\
	&		&	$2$	&	$1.11115$	&	$-0.01031$	&	$-0.01083$	&	$-0.01134$	&	$-0.01184$	\\
	&		&		&	$-0.01235$	&	$-0.01286$	&	$-0.04162$	&				&				\\
	&		&	$3$	&	$1.12790$	&	$-0.01202$	&	$-0.01259$	&	$-0.01316$	&	$-0.01372$	\\
	&		&		&	$-0.01428$	&	$-0.06214$	&				&				&				\\
	&		&	$4$	&	$1.15005$	&	$-0.01429$	&	$-0.01494$	&	$-0.01558$	&	$-0.01621$	\\
	&		&		&	$-0.08903$	&				&				&				&				\\
$2$	&	$6$	&	$0$	&	$1.13843$	&	$-0.02697$	&	$-0.02722$	&	$-0.02753$	&	$-0.02793$	\\
	&		&		&	$-0.02877$	&				&				&				&				\\
	&		&	$1$	&	$1.16664$	&	$-0.03254$	&	$-0.03284$	&	$-0.0332$	&	$-0.06807$	\\
	&		&	$2$	&	$1.20869$	&	$-0.04086$	&	$-0.04122$	&	$-0.12662$	&				\\
	& $10$	&	$0$	&	$1.08942$	&	$-0.00965$	&	$-0.00970$	&	$-0.00976$	&	$-0.00982$	\\
	&		&		&	$-0.00988$	&	$-0.00996$	&	$-0.01005$	&	$-0.01017$	&	$-0.01044$	\\
	&		&	$1$	&	$1.09960$	&	$-0.01076$	&	$-0.01082$	&	$-0.01088$	&	$-0.01094$	\\
	&		&		&	$-0.01102$	&	$-0.01110$	&	$-0.01120$	&	$-0.02288$	&				\\
	&		&	$2$	&	$1.11225$	&	$-0.01215$	&	$-0.01221$	&	$-0.01228$	&	$-0.01235$	\\
	&		&		&	$-0.01243$	&	$-0.01252$	&	$-0.03832$	&				&				\\
	&		&	$3$	&	$1.12846$	&	$-0.01393$	&	$-0.01399$	&	$-0.01407$	&	$-0.01415$	\\
	&		&		&	$-0.01424$	&	$-0.05808$	&				&				&				\\
	&		&	$4$	&	$1.15001$	&	$-0.01629$	&	$-0.01637$	&	$-0.01645$	&	$-0.01655$	\\
	&		&		&	$-0.08436$	&				&				&				&				\\
$3$	&	$6$	&	$0$	&	$1.13871$	&	$-0.02752$	&	$-0.02759$	&	$-0.02768$	&	$-0.0278$	\\
	&		&		&	$-0.02812$	&				&				&				&				\\
	&		&	$1$	&	$1.16665$	&	$-0.03309$	&	$-0.03317$	&	$-0.03327$	&	$-0.06712$	\\
	&		&	$2$	&	$1.20844$	&	$-0.04142$	&	$-0.04152$	&	$-0.12550$	&				\\
	& $10$	&	$0$	&	$1.08980$	&	$-0.00989$	&	$-0.00991$	&	$-0.00992$	&	$-0.00994$	\\
	&		&		&	$-0.00996$	&	$-0.00998$	&	$-0.01001$	&	$-0.01005$	&	$-0.01015$	\\
	&		&	$1$	&	$1.09987$	&	$-0.01101$	&	$-0.01102$	&	$-0.01104$	&	$-0.01106$	\\
	&		&		&	$-0.01108$	&	$-0.01110$	&	$-0.01113$	&	$-0.02244$	&				\\
	&		&	$2$	&	$1.11242$	&	$-0.01240$	&	$-0.01241$	&	$-0.01243$	&	$-0.01245$	\\
	&		&		&	$-0.01247$	&	$-0.01250$	&	$-0.03776$	&				&				\\
	&		&	$3$	&	$1.12854$	&	$-0.01418$	&	$-0.0142$	&	$-0.01422$	&	$-0.01424$	\\
	&		&		&	$-0.01427$	&	$-0.05743$	&				&				&				\\
	&		&	$4$	&	$1.15000$	&	$-0.01656$	&	$-0.01658$	&	$-0.01660$	&	$-0.01663$	\\
	&		&		&	$-0.08365$	&				&				&				&				\\
$4$	&	$6$	&	$0$	&	$1.13881$	&	$-0.02767$	&	$-0.0277$	&	$-0.02773$	&	$-0.02778$	\\
	&		&		&	$-0.02793$	&				&				&				&				\\
	&		&	$1$	&	$1.16666$	&	$-0.03323$	&	$-0.03326$	&	$-0.0333$	&	$-0.06685$	\\
	&		&	$2$	&	$1.20838$	&	$-0.04156$	&	$-0.04160$	&	$-0.12521$	&				\\
	& $10$	&	$0$	&	$1.08991$	&	$-0.00996$	&	$-0.00996$	&	$-0.00997$	&	$-0.00997$	\\
	&		&		&	$-0.00998$	&	$-0.00999$	&	$-0.01000$	&	$-0.01002$	&	$-0.01007$	\\
	&		&	$1$	&	$1.09994$	&	$-0.01107$	&	$-0.01108$	&	$-0.01108$	&	$-0.01109$	\\
	&		&		&	$-0.01110$	&	$-0.01110$	&	$-0.01112$	&	$-0.02231$	&				\\
	&		&	$2$	&	$1.11247$	&	$-0.01246$	&	$-0.01246$	&	$-0.01247$	&	$-0.01248$	\\
	&		&		&	$-0.01249$	&	$-0.0125$	&	$-0.03761$	&				&				\\
	&		&	$3$	&	$1.12856$	&	$-0.01424$	&	$-0.01425$	&	$-0.01426$	&	$-0.01427$	\\
	&		&		&	$-0.01428$	&	$-0.05726$	&				&				&				\\
	&		&	$4$	&	$1.15000$	&	$-0.01662$	&	$-0.01663$	&	$-0.01664$	&	$-0.01665$	\\
	&		&		&	$-0.08346$	&				&				&				&				\\

\hline
\end{longtable}

\setlength{\tabcolsep}{1.57em}
\small{\bf Table 6:} Coefficients for the BLIEs of the scale parameter.
\vspace{-0.2cm}
\small
\begin{longtable}{|r|c|c|r|r|r|r|r|}
\multicolumn{8}{r}{{({\em Continued})}}\\[.7ex]
\endfoot
\endlastfoot
\hline
\multicolumn{1}{|l|}{$~\psi$} & $n$        & $m$          & \multicolumn{5}{c|}{$b'_i$,~$i =1,2,3,\ldots,(n-m)$} \\
\hline
$1$	&	$6$	&	$0$	&	$-0.64800$	&	$0.11675$	&	$0.12276$	&	$0.12885$	&	$0.13536$	\\
	&		&		&	$0.14428$	&				&				&				&				\\
	&		&	$1$	&	$-0.78128$	&	$0.14161$	&	$0.14884$	&	$0.15612$	&	$0.33471$	\\
	&		&	$2$	&	$-0.97992$	&	$0.17906$	&	$0.18811$	&	$0.61276$	&				\\	
	& $10$	&	$0$	&	$-0.68605$	&	$0.06685$	&	$0.06914$	&	$0.07140$	&	$0.07364$	\\
	&		&		&	$0.07588$	&	$0.07816$	&	$0.08056$	&	$0.08324$	&	$0.08717$	\\
	&		&	$1$	&	$-0.7647$	&	$0.07475$	&	$0.07730$	&	$0.07982$	&	$0.08231$	\\
	&		&		&	$0.08481$	&	$0.08735$	&	$0.08998$	&	$0.18839$	&				\\
	&		&	$2$	&	$-0.86267$	&	$0.08467$	&	$0.08756$	&	$0.09040$	&	$0.09321$	\\
	&		&		&	$0.09602$	&	$0.09886$	&	$0.31195$	&				&				\\
	&		&	$3$	&	$-0.98819$	&	$0.09747$	&	$0.10078$	&	$0.10404$	&	$0.10726$	\\
	&		&		&	$0.11047$	&	$0.46816$	&				&				&				\\
	&		&	$4$	&	$-1.15506$	&	$0.11457$	&	$0.11846$	&	$0.12227$	&	$0.12603$	\\
	&		&		&	$0.67373$	&				&				&				&				\\
$2$	&	$6$	&	$0$	&	$-1.48617$	&	$0.29051$	&	$0.29289$	&	$0.29576$	&	$0.29956$	\\
	&		&		&	$0.30744$	&				&				&				&				\\
	&		&	$1$	&	$-1.78762$	&	$0.35003$	&	$0.35289$	&	$0.35631$	&	$0.72840$	\\	
	&	$6$	&	$2$	&	$-2.23759$	&	$0.43902$	&	$0.44258$	&	$1.35598$	&				\\	
	& $10$	&	$0$	&	$-1.59801$	&	$0.17315$	&	$0.17392$	&	$0.17476$	&	$0.17568$	\\
	&		&		&	$0.17671$	&	$0.17790$	&	$0.17933$	&	$0.18125$	&	$0.18533$	\\
	&		&	$1$	&	$-1.77882$	&	$0.19291$	&	$0.19377$	&	$0.19470$	&	$0.19573$	\\
	&		&		&	$0.19687$	&	$0.19818$	&	$0.19976$	&	$0.40688$	&				\\
	&		&	$2$	&	$-2.00375$	&	$0.21754$	&	$0.21851$	&	$0.21956$	&	$0.22071$	\\
	&		&		&	$0.22199$	&	$0.22346$	&	$0.68200$	&				&				\\
	&		&	$3$	&	$-2.29219$	&	$0.24915$	&	$0.25026$	&	$0.25145$	&	$0.25276$	\\
	&		&		&	$0.25422$	&	$1.03435$	&				&				&				\\
	&		&	$4$	&	$-2.67611$	&	$0.29125$	&	$0.29254$	&	$0.29393$	&	$0.29546$	\\
	&		&		&	$1.50294$	&				&				&				&				\\
$3$	&	$6$	&	$0$	&	$-2.34481$	&	$0.46546$	&	$0.46655$	&	$0.46793$	&	$0.46989$	\\
	&		&		&	$0.47498$	&				&				&				&				\\
	&		&	$1$	&	$-2.81661$	&	$0.55947$	&	$0.56077$	&	$0.56241$	&	$1.13396$	\\	
	&		&	$2$	&	$-3.52263$	&	$0.70020$	&	$0.70181$	&	$2.12061$	&				\\	
	& $10$	&	$0$	&	$-2.52877$	&	$0.27877$	&	$0.27910$	&	$0.27948$	&	$0.27990$	\\
	&		&		&	$0.28040$	&	$0.28098$	&	$0.28173$	&	$0.28280$	&	$0.28560$	\\
	&		&	$1$	&	$-2.81202$	&	$0.31010$	&	$0.31047$	&	$0.31089$	&	$0.31136$	\\
	&		&		&	$0.31190$	&	$0.31256$	&	$0.31338$	&	$0.63136$	&				\\
	&		&	$2$	&	$-3.16515$	&	$0.34917$	&	$0.34959$	&	$0.35006$	&	$0.35059$	\\
	&		&		&	$0.35120$	&	$0.35193$	&	$1.06260$	&				&				\\
	&		&	$3$	&	$-3.61863$	&	$0.39936$	&	$0.39984$	&	$0.40037$	&	$0.40098$	\\
	&		&		&	$0.40167$	&	$1.61641$	&				&				&				\\
	&		&	$4$	&	$-4.22284$	&	$0.46623$	&	$0.46679$	&	$0.46742$	&	$0.46812$	\\
	&		&		&	$2.35428$	&				&				&				&				\\
$4$	&	$6$	&	$0$	&	$-3.20022$	&	$0.63805$	&	$0.63863$	&	$0.63938$	&	$0.64047$	\\
	&		&		&	$0.64370$	&				&				&				&				\\
	&		&	$1$	&	$-3.84209$	&	$0.76624$	&	$0.76693$	&	$0.76782$	&	$1.54110$	\\	
	&		&	$2$	&	$-4.80374$	&	$0.95831$	&	$0.95917$	&	$2.88626$	&				\\
	& $10$	&	$0$	&	$-3.45419$	&	$0.38257$	&	$0.38274$	&	$0.38294$	&	$0.38317$	\\
	&		&		&	$0.38343$	&	$0.38376$	&	$0.38418$	&	$0.38479$	&	$0.38662$	\\
	&		&	$1$	&	$-3.83947$	&	$0.42530$	&	$0.42550$	&	$0.42572$	&	$0.42597$	\\
	&		&		&	$0.42626$	&	$0.42662$	&	$0.42708$	&	$0.85702$	&				\\
	&		&	$2$	&	$-4.32042$	&	$0.47866$	&	$0.47887$	&	$0.47912$	&	$0.47940$	\\
	&		&		&	$0.47973$	&	$0.48014$	&	$1.44450$	&				&				\\
	&		&	$3$	&	$-4.93842$	&	$0.54721$	&	$0.54746$	&	$0.54775$	&	$0.54807$	\\
	&		&		&	$0.54844$	&	$2.19948$	&				&				&				\\
	&		&	$4$	&	$-5.76214$	&	$0.63860$	&	$0.63889$	&	$0.63922$	&	$0.63959$	\\
	&		&		&	$3.20583$	&				&				&				&				\\	
\hline
\end{longtable}

\vspace{-0.1cm}
Next, we intend to check the efficiency of the linear estimators suggested in this section. 
The mean squared errors (MSEs) of the BLUE and BLIE  of $\varphi$  may be given by
\[{\rm MSE}(\widehat{\varphi})=\sigma^2 V_1,\]
and 
\begin{align*}
{\rm MSE}(\widetilde{\varphi})= \sigma^{2}\left\lbrace \frac{{\boldsymbol \varphi}^{\text{T}}{\bf{\Sigma}}^{-1}{\boldsymbol \varphi}+1}{({\boldsymbol \varphi}^{\text{T}}{\bf{\Sigma}}^{-1}{\boldsymbol \varphi})(\textbf{1}^{\text{T}}{\bf{\Sigma}}^{-1}\textbf{1})-({\boldsymbol \varphi}^{\text{T}}{\bf{\Sigma}}^{-1}\textbf{1})^2+\textbf{1}^{\text{T}}{\bf{\Sigma}}^{-1}\textbf{1}}\right\rbrace = \sigma^2\left(V_1-\frac{V_3^2}{1+V_2}\right),
\end{align*}
respectively.

The MSEs of the BLUE and BLIE of  $\sigma$ are
\[
{\rm MSE}(\widehat{\sigma})=\sigma^2
V_2,\]
and
\begin{align*}
{\rm MSE}(\widetilde{\sigma})= \sigma^{2}\left\lbrace \frac{\textbf{1}^{\text{T}}{\bf{\Sigma}}^{-1}\textbf{1}}{({\boldsymbol \varphi} ^{\text{T}}{\bf{\Sigma}}^{-1}{\boldsymbol \varphi} )(\textbf{1}^{\text{T}}{\bf{\Sigma}}^{-1}\textbf{1})-({\boldsymbol \varphi} ^{\text{T}}{\bf{\Sigma}}^{-1}\textbf{1})^2+\textbf{1}^{\text{T}}{\bf{\Sigma}}^{-1}\textbf{1}}\right\rbrace = \frac{\sigma^2V_2}{1+V_2},
\end{align*}
respectively.

\vspace{0.2cm}
Thus, the relative efficiency criterion (REC) of the BLIE of  $\varphi$ with respect to the  BLUE of $\varphi$ is given by
\begin{equation}\label{rec1}
{\rm REC}(\widetilde{\varphi},
\widehat{\varphi})=
\frac{V_1}{V_1-\frac{V_3^2}{1+V_2}}.
\end{equation}
Similarly,  the REC of the BLIE of  $\sigma$ w.r.t. the  BLUE of $\sigma$ is given by
\begin{equation}\label{rec2}
{\rm REC}(\widetilde{\sigma},
\widehat{\sigma})=\frac{{\rm MSE}(\widehat{\sigma})}{{\rm MSE}(\widetilde{\sigma})}=
1+V_2.
\end{equation}
It is evident that both RECs are greater than 1 which verifies the superiority of the BLIEs.

\vspace{0.1cm}
We calculated the variances and covariances for the BLUEs of the location and scale parameters in terms of $\sigma^2$. Additionally, we calculated the RECs (\ref{rec1}) and (\ref{rec2}) for Type-II right-censored samples of sizes $n=6$ and 10 with $m$ ranging from 0 to  $[n/2]-1$ and $\psi=1,\cdots,4$. The results are summarized in Table 9.

\vspace{0.2cm}
\setlength{\tabcolsep}{0.8em}
\small{\bf Table 7:} Simulated quantiles of $T_1$ and $T_2$.
\vspace{-0.2cm}
\small
\begin{longtable}{|c|c|c|c|c|c|c|c|c|c|c|}\label{tabcri1}
\endfoot
\endlastfoot
\hline
$\psi$&	$n$&	$m$&		\multicolumn{4}{c|}{$T_1$} 	&\multicolumn{4}{c|}{$T_2$}\\ \hline
&	&	&	0.025&	0.05&	0.95&	0.975&	0.025&	0.05&	0.95&	0.975\\ \hline
1&	6&	0&	-0.8978&	-0.8737&	2.8295&	3.9428&	-1.5398&	-1.3839&	1.8656&	2.3050\\
&	&	1&	-0.8807&	-0.8568&	2.9112&	4.1003&	-1.4721&	-1.3298&	1.8152&	2.3110\\
&	&	2&	-0.8521&	-0.8288&	3.5163&	5.3046&	-1.4026&	-1.2924&	1.8865&	2.4439\\
&	10&	0&	-0.9321&	-0.9082&	2.4709&	3.5555&	-1.6687&	-1.4476&	1.8083&	2.2341\\
&	&	1&	-0.9250&	-0.8990&	2.4891&	3.6256&	-1.6572&	-1.4439&	1.8288&	2.3033\\
&	&	2&	-0.9173&	-0.8950&	2.5101&	3.5760&	-1.6216&	-1.4252&	1.8541&	2.2587\\
&	&	3&	-0.9099&	-0.8845&	2.6774&	3.7372&	-1.5690&	-1.3974&	1.8148&	2.2806\\
&	&	4&	-0.8978&	-0.8746&	2.8998&	4.1416&	-1.5393&	-1.3829&	1.8490&	2.3085\\
2&	6&	0&	-0.8897&	-0.8663&	2.7817&	3.9359&	-1.5009&	-1.3672&	1.7675&	2.2691\\
&	&	1&	-0.8726&	-0.8489&	3.0881&	4.5173&	-1.4785&	-1.3157&	1.8894&	2.4233\\
&	&	2&	-0.8440&	-0.8227&	3.6865&	5.5509&	-1.3804&	-1.2664&	1.8961&	2.4077\\
&	10&	0&	-0.9260&	-0.9012&	2.3966&	3.3919&	-1.6343&	-1.4574&	1.7925&	2.2208\\
&	&	1&	-0.9197&	-0.8964&	2.6083&	3.5056&	-1.6053&	-1.4251&	1.8601&	2.3783\\
&	&	2&	-0.9137&	-0.8867&	2.5283&	3.4814&	-1.5701&	-1.4034&	1.8431&	2.2891\\
&	&	3&	-0.9039&	-0.8822&	2.6312&	3.7844&	-1.5539&	-1.3877&	1.8838&	2.3184\\
&	&	4&	-0.8917&	-0.8693&	2.8245&	4.0445&	-1.4962&	-1.3358&	1.9245&	2.3871\\
3&	6&	0&	-0.8925&	-0.8639&	2.7511&	4.0257&	-1.5196&	-1.3476&	1.8517&	2.3827\\
&	&	1&	-0.8723&	-0.8503&	3.1442&	4.6298&	-1.4449&	-1.3173&	1.8643&	2.3721\\
&	&	2&	-0.8436&	-0.8197&	3.6674&	5.5449&	-1.3846&	-1.2760&	1.8893&	2.4688\\
&	10&	0&	-0.9263&	-0.9023&	2.3569&	3.2632&	-1.6431&	-1.4281&	1.8186&	2.2366\\
&	&	1&	-0.9199&	-0.8936&	2.4461&	3.3830&	-1.6130&	-1.4119&	1.8098&	2.2553\\
&	&	2&	-0.9090&	-0.8865&	2.4581&	3.5173&	-1.5609&	-1.3930&	1.8141&	2.2439\\
&	&	3&	-0.9022&	-0.8780&	2.6821&	3.8591&	-1.5594&	-1.3945&	1.8051&	2.2466\\
&	&	4&	-0.8886&	-0.8637&	2.9251&	4.1946&	-1.4937&	-1.3422&	1.8371&	2.3322\\
4&	6&	0&	-0.8924&	-0.8671&	2.8063&	4.2672&	-1.5281&	-1.3619&	1.8677&	2.3828\\
&	&	1&	-0.8722&	-0.8487&	3.0416&	4.3763&	-1.4421&	-1.3071&	1.9235&	2.3853\\
&	&	2&	-0.8423&	-0.8199&	3.7080&	5.6595&	-1.3789&	-1.2573&	1.8838&	2.4782\\
&	10&	0&	-0.9266&	-0.9012&	2.3821&	3.3501&	-1.6292&	-1.4180&	1.7917&	2.2129\\
&	&	1&	-0.9197&	-0.8945&	2.4726&	3.4174&	-1.6321&	-1.4259&	1.8601&	2.2985\\
&	&	2&	-0.9107&	-0.8869&	2.5691&	3.5821&	-1.5857&	-1.3812&	1.8747&	2.2729\\
&	&	3&	-0.9025&	-0.8795&	2.7359&	3.7134&	-1.5375&	-1.3813&	1.8365&	2.3429\\
&	&	4&	-0.8885&	-0.8644&	2.8087&	4.1154&	-1.5091&	-1.3436&	1.8384&	2.3835\\
 \hline
\end{longtable}

\setlength{\tabcolsep}{0.8em}
\small{\bf Table 8:} Simulated quantiles of $T_3$ and $T_4$.
\vspace{-0.2cm}
\small
\begin{longtable}{|c|c|c|c|c|c|c|c|c|c|c|}\label{tabcri2}
\endfoot
\endlastfoot
\hline
$\psi$&	$n$&	$m$&		\multicolumn{4}{c|}{$T_3$} 	&\multicolumn{4}{c|}{$T_4$}\\ \hline
&	&	&	0.025&	0.05&	0.95&	0.975&	0.025&	0.05&	0.95&	0.975\\ \hline
1&	6&	0&	-0.9080&	-0.8786&	3.6233&	4.9767&	-1.9699&	-1.8141&	1.4354&	1.8748\\
&	&	1&	-0.8991&	-0.8685&	3.9549&	5.4769&	-1.9556&	-1.8133&	1.3317&	1.8275\\
&	&	2&	-0.8854&	-0.8528&	5.2013&	7.6929&	-1.9642&	-1.8540&	1.3248&	1.8823\\
&	10&	0&	-0.9329&	-0.9063&	2.8540&	4.0611&	-1.9889&	-1.7678&	1.4881&	1.9139\\
&	&	1&	-0.9271&	-0.8977&	2.9274&	4.2105&	-1.9978&	-1.7845&	1.4882&	1.9627\\
&	&	2&	-0.9214&	-0.8957&	3.0208&	4.2469&	-1.9870&	-1.7906&	1.4887&	1.8933\\
&	&	3&	-0.9175&	-0.8875&	3.3126&	4.5623&	-1.9651&	-1.7935&	1.4187&	1.8846\\
&	&	4&	-0.9106&	-0.8822&	3.7262&	5.2424&	-1.9749&	-1.8184&	1.4134&	1.8730\\
2&	6&	0&	-0.9062&	-0.8774&	3.6002&	5.0169&	-1.9437&	-1.8100&	1.3247&	1.8263\\
&	&	1&	-0.8977&	-0.8671&	4.2248&	6.0733&	-1.9747&	-1.8119&	1.3932&	1.9271\\
&	&	2&	-0.8841&	-0.8540&	5.4973&	8.1232&	-1.9544&	-1.8404&	1.3221&	1.8337\\
&	10&	0&	-0.9311&	-0.9033&	2.7885&	3.9027&	-1.9644&	-1.7874&	1.4625&	1.8908\\
&	&	1&	-0.9264&	-0.8999&	3.0818&	4.1011&	-1.9559&	-1.7756&	1.5095&	2.0277\\
&	&	2&	-0.9228&	-0.8915&	3.0621&	4.1656&	-1.9453&	-1.7786&	1.4679&	1.9139\\
&	&	3&	-0.9163&	-0.8906&	3.2806&	4.6497&	-1.9596&	-1.7934&	1.4780&	1.9127\\
&	&	4&	-0.9095&	-0.8820&	3.6599&	5.1599&	-1.9411&	-1.7807&	1.4796&	1.9422\\
3&	6&	0&	-0.9112&	-0.8760&	3.5713&	5.1393&	-1.9653&	-1.7933&	1.4060&	1.9370\\
&	&	1&	-0.8989&	-0.8704&	4.3071&	6.2327&	-1.9437&	-1.8160&	1.3656&	1.8734\\
&	&	2&	-0.8853&	-0.8515&	5.4818&	8.1318&	-1.9609&	-1.8523&	1.3130&	1.8925\\
&	10&	0&	-0.9322&	-0.9054&	2.7482&	3.7642&	-1.9753&	-1.7602&	1.4864&	1.9044\\
&	&	1&	-0.9275&	-0.8975&	2.9016&	3.9675&	-1.9656&	-1.7645&	1.4572&	1.9027\\
&	&	2&	-0.9184&	-0.8923&	2.9850&	4.2127&	-1.9379&	-1.7701&	1.4370&	1.8669\\
&	&	3&	-0.9153&	-0.8866&	3.3453&	4.7445&	-1.9669&	-1.8019&	1.3976&	1.8391\\
&	&	4&	-0.9067&	-0.8760&	3.7883&	5.3511&	-1.9402&	-1.7887&	1.3906&	1.8857\\
4&	6&	0&	-0.9115&	-0.8803&	3.6420&	5.4406&	-1.9747&	-1.8085&	1.4211&	1.9363\\
&	&	1&	-0.8992&	-0.8688&	4.1768&	5.9079&	-1.9415&	-1.8065&	1.4240&	1.8858\\
&	&	2&	-0.8837&	-0.8522&	5.5424&	8.2984&	-1.9558&	-1.8342&	1.3069&	1.9013\\
&	10&	0&	-0.9329&	-0.9044&	2.7778&	3.8633&	-1.9620&	-1.7508&	1.4588&	1.8801\\
&	&	1&	-0.9275&	-0.8988&	2.9331&	4.0083&	-1.9852&	-1.7790&	1.5070&	1.9454\\
&	&	2&	-0.9206&	-0.8930&	3.1149&	4.2895&	-1.9633&	-1.7588&	1.4971&	1.8953\\
&	&	3&	-0.9159&	-0.8887&	3.4105&	4.5729&	-1.9454&	-1.7892&	1.4286&	1.9350\\
&	&	4&	-0.9068&	-0.8772&	3.6464&	5.2555&	-1.9561&	-1.7905&	1.3915&	1.9366\\
 \hline
\end{longtable}

\setlength{\tabcolsep}{1.33em}
\small{\bf Table 9:} Variances and covariances for the BLUEs of the location and scale parameters in terms of $\sigma^2$ and the RECs of the BLIEs  w.r.t. the BLUEs.
\vspace{-0.2cm}
\small
\begin{longtable}{|c|c|c|c|c|c|c|c|}
\endfoot
\endlastfoot
\hline
$~\psi$ & $n$        & $m$          & $V_1$	& $V_2$	& $V_3$ &${\rm REC}(\widetilde{\varphi},
\widehat{\varphi})$&${\rm REC}(\widetilde{\sigma},
\widehat{\sigma})$\\
\hline

$1$	&	$6$	&	$0$	&	$0.05615$	&	$0.18505$	&	$-0.04238$	&	$1.02774$	&	$1.18505$	\\
	&		&	$1$	&	$0.05857$	&	$0.23378$	&	$-0.05325$	&	$1.04084$	&	$1.23378$	\\
	&		&	$2$	&	$0.06256$	&	$0.31541$	&	$-0.07129$	&	$1.06582$	&	$1.31541$	\\
	& $10$	&	$0$	&	$0.01908$	&	$0.10250$	&	$-0.01428$	&	$1.00979$	&	$1.10250$	\\
	&		&	$1$	&	$0.01933$	&	$0.11602$	&	$-0.01612$	&	$1.01219$	&	$1.11602$	\\
	&		&	$2$	&	$0.01964$	&	$0.13350$	&	$-0.01847$	&	$1.01556$	&	$1.13349$	\\
	&		&	$3$	&	$0.02006$	&	$0.15688$	&	$-0.02161$	&	$1.02052$	&	$1.15688$	\\
	&		&	$4$	&	$0.02065$	&	$0.18970$	&	$-0.02598$	&	$1.02826$	&	$1.18970$	\\
$2$	&	$6$	&	$0$	&	$0.01045$	&	$0.19612$	&	$-0.01852$	&	$1.02820$	&	$1.19612$	\\
	&		&	$1$	&	$0.01089$	&	$0.24622$	&	$-0.02321$	&	$1.04134$	&	$1.24622$	\\
	&		&	$2$	&	$0.01162$	&	$0.32949$	&	$-0.03101$	&	$1.06637$	&	$1.32949$	\\
	& $10$	&	$0$	&	$0.00350$	&	$0.10892$	&	$-0.00618$	&	$1.00996$	&	$1.10892$	\\
	&		&	$1$	&	$0.00354$	&	$0.12288$	&	$-0.00697$	&	$1.01237$	&	$1.12288$	\\
	&		&	$2$	&	$0.00360$	&	$0.14078$	&	$-0.00798$	&	$1.01576$	&	$1.14078$	\\
	&		&	$3$	&	$0.00367$	&	$0.16461$	&	$-0.00932$	&	$1.02072$	&	$1.16461$	\\
	&		&	$4$	&	$0.00378$	&	$0.19794$	&	$-0.01119$	&	$1.02847$	&	$1.19794$	\\
$3$	&	$6$	&	$0$	&	$0.00420$	&	$0.19864$	&	$-0.01180$	&	$1.02843$	&	$1.19864$	\\
	&		&	$1$	&	$0.00438$	&	$0.24874$	&	$-0.01477$	&	$1.04155$	&	$1.24874$	\\
	&		&	$2$	&	$0.00467$	&	$0.33210$	&	$-0.01971$	&	$1.06656$	&	$1.33210$	\\
	& $10$	&	$0$	&	$0.00140$	&	$0.11035$	&	$-0.00394$	&	$1.01005$	&	$1.11035$	\\
	&		&	$1$	&	$0.00142$	&	$0.12429$	&	$-0.00443$	&	$1.01245$	&	$1.12429$	\\
	&		&	$2$	&	$0.00144$	&	$0.14218$	&	$-0.00507$	&	$1.01583$	&	$1.14218$	\\
	&		&	$3$	&	$0.00147$	&	$0.16601$	&	$-0.00591$	&	$1.02079$	&	$1.16601$	\\
	&		&	$4$	&	$0.00152$	&	$0.19936$	&	$-0.00710$	&	$1.02854$	&	$1.19936$	\\
$4$	&	$6$	&	$0$	&	$0.00226$	&	$0.19942$	&	$-0.00866$	&	$1.02851$	&	$1.19940$	\\
	&		&	$1$	&	$0.00235$	&	$0.24948$	&	$-0.01084$	&	$1.04161$	&	$1.24947$	\\
	&		&	$2$	&	$0.00251$	&	$0.33283$	&	$-0.01445$	&	$1.06662$	&	$1.33283$	\\
	& $10$	&	$0$	&	$0.00075$	&	$0.11078$	&	$-0.00289$	&	$1.01008$	&	$1.11078$	\\
	&		&	$1$	&	$0.00076$	&	$0.12470$	&	$-0.00325$	&	$1.01248$	&	$1.12470$	\\
	&		&	$2$	&	$0.00078$	&	$0.14258$	&	$-0.00372$	&	$1.01586$	&	$1.14258$	\\
	&		&	$3$	&	$0.00079$	&	$0.16640$	&	$-0.00434$	&	$1.02082$	&	$1.16640$	\\
	&		&	$4$	&	$0.00081$	&	$0.19974$	&	$-0.00521$	&	$1.02856$	&	$1.19974$	\\
 \hline
\end{longtable}

\section{Linear Prediction}
Let $X_{1:n} \leq X_{2:n} \leq ... \leq X_{n-m:n},~m=1,..., n-1,$ represent a Type-II right censored sample from $XL(\varphi,\sigma\psi)$. Furthermore, define $\text{Z}_{r:n} =\dfrac{X_{r:n}-\varphi}{\sigma}$, where $\text{E}(\text{Z}_{r:n})=\varphi^{(1)}_{r:n},~1 \leq  r \leq n-m,$ and $\text{Cov}(\text{Z}_{r:n}, \text{Z}_{s:n})=\sigma_{r,s:n}=\varphi^{(1,1)}_{r,s:n}-\varphi^{(1)}_{r:n}\varphi^{(1)}_{s:n}, 1 \leq r < s \leq n-m$. 

\vspace{0.2cm}
Suppose  $\text{Z}_{q:n}=\dfrac{X_{q:n}-\varphi}{\sigma}$ for $n-m+1\leq q \leq n$. So, the best linear unbiased predictor (BLUP) of $X_{q:n}$ can be expressed as
(see \cite{KN})
\[
\widehat{X}_{q:n}=\widehat{\varphi}+\widehat{\sigma}\varphi_{q:n}+{\boldsymbol \textomega}^{\text{T}}{\bf{\Sigma}}^{-1}\left({\boldsymbol X}-\widehat{\varphi}{\boldsymbol 1}-
\widehat{\sigma}{\boldsymbol \varphi}\right),
\]
where   $\varphi_{q:n}$ is the mean of $\text{Z}_{q:n}$, 
$\widehat{\varphi}$ and $\widehat{\sigma}$ are the BLUEs of the location and scale parameters, respectively, 
and
\[
{\boldsymbol \textomega}^{\text{T}}=\left(\text{Cov}(\text{Z}_{1:n},\text{Z}_{q:n}),\cdots, \text{Cov}(\text{Z}_{n-m:n},\text{Z}_{q:n})\right).
\]

The mean squared prediction error (MSPE) of
$\widehat{X}_{q:n}$ is  (see for example, Burkshat \cite{Burk})
\begin{eqnarray*}
{\rm MSPE}(\widehat{X}_{q:n})&=&\text{E}\left[\left(\widehat{X}_{q:n}-X_{q:n}\right)^2\right]\\&=&
\sigma^2\bigg[\left(1-{\boldsymbol  \textomega}^{\text{T}} {\bf{\Sigma}}^{-1}\mathbf{1}\right)^2V_1+\left(\varphi_{q:n}-{\boldsymbol  \textomega}^{\text{T}} {\bf{\Sigma}}^{-1}
{\boldsymbol  \varphi}\right)^2V_2-{\boldsymbol  \textomega}^{\text{T}} {\bf{\Sigma}}^{-1}{\boldsymbol  \textomega}\\&&+2\left(1-{\boldsymbol  \textomega}^{\text{T}} {\bf{\Sigma}}^{-1}\mathbf{1}\right)\left(\varphi_{q:n}-{\boldsymbol  \textomega}^{\text{T}} {\bf{\Sigma}}^{-1}{\boldsymbol  \varphi}\right)V_3+\text{Var}(\text{Z}_{q:n})\bigg].
\end{eqnarray*}

According to the findings of Mann \cite{Man}, the best linear invariant predictor (BLIP) of $X_{q:n}$ can be expressed as
\begin{equation*}
\widetilde{X}_{q:n}=\widetilde{\varphi}+\widetilde{\sigma}\varphi_{q:n}+{\boldsymbol \textomega}^{\text{T}}{\bf{\Sigma}}^{-1}\left({\boldsymbol X}-\widetilde{\varphi}{\boldsymbol 1}-
\widetilde{\sigma}{\boldsymbol \varphi}\right)=\widehat{X}_{q:n}-\left(\frac{V_4}{1+V_2}\right)\widehat{\sigma},
\end{equation*}
where $\widetilde{\varphi}$ and $\widetilde{\sigma}$ are the BLIEs of the location and scale parameters, respectively, and
\[V_4=(1-{\boldsymbol  \textomega}^{\text{T}} {\bf{\Sigma}}^{-1} {\boldsymbol 1} )V_3+\left(\varphi_{q:n} -{\boldsymbol  \textomega}^{\text{T}}
 {\bf{\Sigma}}^{-1}{\boldsymbol  \varphi}\right)V_2.\]
The MSPE of $\widetilde{X}_{q:n}$ is given by (see for example, Burkshat \cite{Burk})
\begin{eqnarray*}
{\rm MSPE}(\widetilde{X}_{q:n})\hspace*{-.2cm}&=&\hspace*{-.2cm}\text{E}\left[\left(\widetilde{X}_{q:n}-X_{q:n}\right)^2\right]\\\hspace*{-.2cm}&=&\hspace*{-.2cm}
\sigma^2\bigg{[}\left(1-{\boldsymbol  \textomega}^{\text{T}} {\bf{\Sigma}}^{-1}\mathbf{1}\right)^2
\left(\frac{V_1}{1+V_2}+\frac{1}{\Delta}\right)
 +
\left(\varphi_{q:n}-{\boldsymbol  \textomega}^{\text{T}} {\bf{\Sigma}}^{-1}{\boldsymbol  \varphi}\right)^2 \frac{V_2}{1+V_2}
-{\boldsymbol  \textomega}^{\text{T}} {\bf{\Sigma}}^{-1}{\boldsymbol  \textomega}
\\&&\quad\quad +2\,
\left(1-{\boldsymbol  \textomega}^{\text{T}} {\bf{\Sigma}}^{-1}\mathbf{1}\right)\left(\varphi_{q:n}-{\boldsymbol  \textomega}^{\text{T}} {\bf{\Sigma}}^{-1}{\boldsymbol  \varphi}\right)
\frac{V_3}{1+V_2}
+\text{Var}(\text{Z}_{q:n})\bigg{]},
\end{eqnarray*}
where
\begin{eqnarray*}
\Delta=({\boldsymbol \varphi} ^{\text{T}}{\bf{\Sigma}}^{-1}{\boldsymbol \varphi} )(\textbf{1}^{\text{T}}{\bf{\Sigma}}^{-1}\textbf{1})-({\boldsymbol \varphi} ^{\text{T}}{\bf{\Sigma}}^{-1}\textbf{1})^2+\textbf{1}^{\text{T}}{\bf{\Sigma}}^{-1}\textbf{1}.
\end{eqnarray*}
To compare the BLUP and BLIP, we consider the REC of $\widetilde{X}_{q:n}$ w.r.t.  $\widehat{X}_{q:n}$, which is given by
\[{\rm REC}(\widetilde{X}_{q:n},\widehat{X}_{q:n})=\frac{{\rm MSPE}(\widehat{X}_{q:n})}{{\rm MSPE}(\widetilde{X}_{q:n})}.\]

Table 10 displays the somputed values of ${\rm REC}(\widetilde{X}_{q:n},\widehat{X}_{q:n})$ and coefficients $V_4$ for $n=6$ and 10 with $m$ ranging from 1
to $([n/2]-1$) and $\psi=1,\cdots,4$. Table 8 verifies the superiority of the  BLIPs over the BLUPs for the cases that are considered.

\vspace{0.2cm}
We now turn our attention to the construction of prediction intervals (PIs) for the order statistic for ${X}_{q:n}$. 
These PIs can be formulated using specific pivotal quantities (see for example Balakrishnan and Chan \cite{Bala}):
\[
T^*_1 = \frac{{X}_{q:n} - X_{(n-m):n}}{\hat\sigma},
\]
and
\[
T^*_2 = \frac{{X}_{q:n} -  X_{(n-m):n}}{\widetilde{\sigma}}.
\]
To establish these PIs, it is essential to determine the percentage points of $T^*_1$ and $T^*_2$. 
The simulated percentage points for these pivotal quantities are presented in Table 11, calculated from a Monte Carlo simulation comprising 10,000 iterations, with different values of $n$ and $\psi$. 

\vspace{0.20cm}
By employing the pivotal quantity $T^*_1$, we can express a $100(1 - \Upsilon)\%$ PI for ${X}_{q:n}$ as follows:
\[
\text{P}\left(  X_{(n-m):n} + \hat\sigma T^*_1(\Upsilon/2) \leq {X}_{q:n} \leq  X_{(n-m):n} + \hat\sigma T^*_1(1 - \Upsilon/2) \right) = 1 - \Upsilon,
\]
where $T^*_1(\tau)$ denotes the left percentage point of $T^*_1$ at $\tau$, i.e., $\text{P}(T^*_1 < T^*_1(\tau)) = \tau$.

\vspace{0.20cm}
Similarly, by employing the pivotal quantity $T^*_2$, we can express a $100(1 - \Upsilon)\%$ PI for ${X}_{q:n}$ as follows:
\[
\text{P}\left(  X_{(n-m):n} + \tilde\sigma T^*_1(\Upsilon/2) \leq {X}_{q:n} \leq  X_{(n-m):n} + \tilde\sigma T^*_1(1 - \Upsilon/2) \right) = 1 - \Upsilon,
\]
where $T^*_2(\tau)$ denotes the left percentage point of $T^*_2$ at $\tau$, i.e., $\text{P}(T^*_2 < T^*_2(\tau)) = \tau$.

\newpage
\small{\bf Table 10:} Coefficients $V_4$ and the RECs of BLIPs  w.r.t. the  BLUPs.
\vspace{-0.3cm}
\setlength{\tabcolsep}{0.76em}
\small
\begin{longtable}{|r|c|c|r|r|r|r|r|r|r|r|}
\multicolumn{8}{r}{{({\em Continued})}}\\[.9ex]
\endfoot
\endlastfoot
\hline
\multicolumn{1}{|l|}{$\psi$} & $n$     & $m$     & \multicolumn{8}{c|}{$(n-m+1)\leq q\leq n$} \\
\multicolumn{1}{|l|}{} &      &      & \multicolumn{4}{c|}{$V_4$}& \multicolumn{4}{c|}{RECs}\\
\hline
$1$	&	$6$	&	$1$ &	$0.28499$	&	&	&	&	$1.04112$	&	&	&	\\
	&		&	$2$	&	$0.19768$	&	$0.57698$	&	&	&	$1.06617$	&	$1.10296$	&	&	\\
	& $10$	&	$1$	&	$0.14063$	&	&	&	&	$1.01226$	&	&	&	\\
$		$	&	$		$	&	$	2	$	&	$	0.08314	$	&	$	0.24312	$	&	$		$	&	$		$	&	$	1.01566	$	&	$	1.02592	$	&	$		$	&	$		$	\\
$		$	&	$		$	&	$	3	$	&	$	0.06613	$	&	$	0.16277	$	&	$	0.34927	$	&	$		$	&	$	1.02063	$	&	$	1.03558	$	&	$	1.0432	$	&	$		$	\\
$		$	&	$		$	&	$	4	$	&	$	0.06059	$	&	$	0.13974	$	&	$	0.25561	$	&	$	0.47973	$	&	$	1.02837	$	&	$	1.04889	$	&	$	1.06281	$	&	$	1.06737	$	\\
$	2	$	&	$	6	$	&	$	1	$	&	$	0.13624	$	&	$		$	&	$		$	&	$		$	&	$	1.04189	$	&	$		$	&	$		$	&	$		$	\\
$		$	&	$		$	&	$	2	$	&	$	0.09174	$	&	$	0.27331	$	&	$		$	&	$		$	&	$	1.06682	$	&	$	1.10406	$	&	$		$	&	$		$	\\
$		$	&	$	10	$	&	$	1	$	&	$	0.06795	$	&	$		$	&	$		$	&	$		$	&	$	1.01259	$	&	$		$	&	$		$	&	$		$	\\
$		$	&	$		$	&	$	2	$	&	$	0.03918	$	&	$	0.11673	$	&	$		$	&	$		$	&	$	1.01594	$	&	$	1.02645	$	&	$		$	&	$		$	\\
$		$	&	$		$	&	$	3	$	&	$	0.03063	$	&	$	0.0763	$	&	$	0.16674	$	&	$		$	&	$	1.02089	$	&	$	1.03607	$	&	$	1.04383	$	&	$		$	\\
$		$	&	$		$	&	$	4	$	&	$	0.02767	$	&	$	0.0644	$	&	$	0.11919	$	&	$	0.22773	$	&	$	1.02862	$	&	$	1.04937	$	&	$	1.06344	$	&	$	1.06798	$	\\
$	3	$	&	$	6	$	&	$	1	$	&	$	0.088	$	&	$		$	&	$		$	&	$		$	&	$	1.0418	$	&	$		$	&	$		$	&	$		$	\\
$		$	&	$		$	&	$	2	$	&	$	0.05887	$	&	$	0.17618	$	&	$		$	&	$		$	&	$	1.06675	$	&	$	1.10377	$	&	$		$	&	$		$	\\
$		$	&	$	10	$	&	$	1	$	&	$	0.04397	$	&	$		$	&	$		$	&	$		$	&	$	1.01256	$	&	$		$	&	$		$	&	$		$	\\
$		$	&	$		$	&	$	2	$	&	$	0.0252	$	&	$	0.07543	$	&	$		$	&	$		$	&	$	1.01591	$	&	$	1.02636	$	&	$		$	&	$		$	\\
$		$	&	$		$	&	$	3	$	&	$	0.01963	$	&	$	0.04903	$	&	$	0.10761	$	&	$		$	&	$	1.02086	$	&	$	1.036	$	&	$	1.04365	$	&	$		$	\\
$		$	&	$		$	&	$	4	$	&	$	0.01769	$	&	$	0.04125	$	&	$	0.07652	$	&	$	0.14681	$	&	$	1.0286	$	&	$	1.04931	$	&	$	1.06331	$	&	$	1.06767	$	\\
$	4	$	&	$	6	$	&	$	1	$	&	$	0.06484	$	&	$		$	&	$		$	&	$		$	&	$	1.04174	$	&	$		$	&	$		$	&	$		$	\\
$		$	&	$		$	&	$	2	$	&	$	0.04329	$	&	$	0.12973	$	&	$		$	&	$		$	&	$	1.06671	$	&	$	1.10361	$	&	$		$	&	$		$	\\
$		$	&	$	10	$	&	$	1	$	&	$	0.03241	$	&	$		$	&	$		$	&	$		$	&	$	1.01253	$	&	$		$	&	$		$	&	$		$	\\
$		$	&	$		$	&	$	2	$	&	$	0.01854	$	&	$	0.05558	$	&	$		$	&	$		$	&	$	1.01589	$	&	$	1.0263	$	&	$		$	&	$		$	\\
$		$	&	$		$	&	$	3	$	&	$	0.01443	$	&	$	0.03606	$	&	$	0.07926	$	&	$		$	&	$	1.02085	$	&	$	1.03596	$	&	$	1.04356	$	&	$		$	\\
$		$	&	$		$	&	$	4	$	&	$	0.013	$	&	$	0.03031	$	&	$	0.05627	$	&	$	0.10811	$	&	$	1.02858	$	&	$	1.04928	$	&	$	1.06324	$	&	$	1.06753	$	\\

\hline
\end{longtable}

\setlength{\tabcolsep}{0.78em}
\small{\bf Table 11:} Simulated quantiles of $T^\ast_1$ and $T^\ast_2$.
\vspace{-0.2cm}
\small
\begin{longtable}{|c|c|c|c|c|c|c|c|c|c|c|c|}\label{tabcri3}
\endfoot
\endlastfoot
\hline
$\psi$&	$n$&	$m$&	$q$&		\multicolumn{4}{c|}{$T^\ast_1$} 	&\multicolumn{4}{c|}{$T^\ast_2$}\\ \hline
1&6&1&6&0.0312&0.0607&4.9687&6.6338&0.0312&0.0607&4.9687&6.6338\\
&&2&5&0.0167&0.0325&3.1080&4.4073&0.0167&0.0325&3.1080&4.4073\\
&&&6&0.2009&0.2892&8.0494&11.3947&0.2009&0.2892&8.0494&11.3947\\
&10&1&10&0.0296&0.0572&4.1384&5.2123&0.0296&0.0572&4.1384&5.2123\\
&&2&9&0.0146&0.0299&2.2299&2.8513&0.0146&0.0299&2.2299&2.8513\\
&&&10&0.1977&0.2941&5.5721&6.9617&0.1977&0.2941&5.5721&6.9617\\
&&3&8&0.0113&0.0206&1.5785&1.9908&0.0113&0.0206&1.5785&1.9908\\
&&&9&0.1086&0.1586&3.1891&4.0463&0.1086&0.1586&3.1891&4.0463\\
&&&10&0.3741&0.5194&6.4680&8.0829&0.3741&0.5194&6.4680&8.0829\\
&&4&7&0.0075&0.0157&1.2673&1.6964&0.0075&0.0157&1.2673&1.6964\\
&&&8&0.0821&0.1201&2.4486&3.1193&0.0821&0.1201&2.4486&3.1193\\
&&&9&0.2342&0.3170&4.2819&5.4436&0.2342&0.3170&4.2819&5.4436\\
&&&10&0.5415&0.6866&7.6615&9.9841&0.5415&0.6866&7.6615&9.9841\\
2&6&1&6&0.0171&0.0312&2.3784&3.1912&0.0171&0.0312&2.3784&3.1912\\
&&2&5&0.0068&0.0142&1.3616&1.9148&0.0068&0.0142&1.3616&1.9148\\
&&&6&0.0870&0.1252&3.7318&5.0208&0.0870&0.1252&3.7318&5.0208\\
&10&1&10&0.0134&0.0270&2.0137&2.6044&0.0134&0.0270&2.0137&2.6044\\
&&2&9&0.0068&0.0138&1.0655&1.3676&0.0068&0.0138&1.0655&1.3676\\
&&&10&0.0918&0.1376&2.5513&3.1611&0.0918&0.1376&2.5513&3.1611\\
&&3&8&0.0053&0.0104&0.7227&0.9525&0.0053&0.0104&0.7227&0.9525\\
&&&9&0.0513&0.0764&1.4957&1.8966&0.0513&0.0764&1.4957&1.8966\\
&&&10&0.1753&0.2309&3.0790&3.9131&0.1753&0.2309&3.0790&3.9131\\
&&4&7&0.0038&0.0075&0.5666&0.7532&0.0038&0.0075&0.5666&0.7532\\
&&&8&0.0380&0.0554&1.1184&1.4284&0.0380&0.0554&1.1184&1.4284\\
&&&9&0.1012&0.1418&1.9145&2.4569&0.1012&0.1418&1.9145&2.4569\\
&&&10&0.2412&0.3169&3.5842&4.4294&0.2412&0.3169&3.5842&4.4294\\
3&6&1&6&0.0081&0.0187&1.5986&2.1286&0.0081&0.0187&1.5986&2.1286\\
&&2&5&0.0043&0.0093&0.9151&1.2752&0.0043&0.0093&0.9151&1.2752\\
&&&6&0.0556&0.0838&2.5309&3.4847&0.0556&0.0838&2.5309&3.4847\\
&10&1&10&0.0089&0.0185&1.2697&1.6020&0.0089&0.0185&1.2697&1.6020\\
&&2&9&0.0048&0.0096&0.6572&0.8626&0.0048&0.0096&0.6572&0.8626\\
&&&10&0.0578&0.0851&1.6583&2.1177&0.0578&0.0851&1.6583&2.1177\\
&&3&8&0.0031&0.0066&0.4591&0.6062&0.0031&0.0066&0.4591&0.6062\\
&&&9&0.0336&0.0491&0.9747&1.2620&0.0336&0.0491&0.9747&1.2620\\
&&&10&0.1079&0.1481&2.0206&2.5351&0.1079&0.1481&2.0206&2.5351\\
&&4&7&0.0023&0.0048&0.3673&0.5027&0.0023&0.0048&0.3673&0.5027\\
&&&8&0.0240&0.0346&0.7319&0.9273&0.0240&0.0346&0.7319&0.9273\\
&&&9&0.0661&0.0891&1.2601&1.5855&0.0661&0.0891&1.2601&1.5855\\
&&&10&0.1535&0.1997&2.3646&2.9617&0.1535&0.1997&2.3646&2.9617\\
4&6&1&6&0.0071&0.0143&1.1443&1.6182&0.0071&0.0143&1.1443&1.6182\\
&&2&5&0.0035&0.0072&0.6715&0.9427&0.0035&0.0072&0.6715&0.9427\\
&&&6&0.0410&0.0617&1.7516&2.4405&0.0410&0.0617&1.7516&2.4405\\
&10&1&10&0.0064&0.0130&0.9314&1.2081&0.0064&0.0130&0.9314&1.2081\\
&&2&9&0.0035&0.0070&0.4874&0.6370&0.0035&0.0070&0.4874&0.6370\\
&&&10&0.0409&0.0621&1.2365&1.5721&0.0409&0.0621&1.2365&1.5721\\
&&3&8&0.0021&0.0043&0.3262&0.4301&0.0021&0.0043&0.3262&0.4301\\
&&&9&0.0229&0.0346&0.7195&0.9092&0.0229&0.0346&0.7195&0.9092\\
&&&10&0.0841&0.1124&1.4966&1.8947&0.0841&0.1124&1.4966&1.8947\\
&&4&7&0.0016&0.0035&0.2679&0.3548&0.0016&0.0035&0.2679&0.3548\\
&&&8&0.0172&0.0256&0.5232&0.6678&0.0172&0.0256&0.5232&0.6678\\
&&&9&0.0483&0.0671&0.9351&1.1675&0.0483&0.0671&0.9351&1.1675\\
&&&10&0.1144&0.1475&1.7556&2.2218&0.1144&0.1475&1.7556&2.2218\\
 \hline
\end{longtable}

\section{A Simulation Study}
Here, we conduct a  simulation to evaluate the suggested estimators and predictors, aiming to determine whether the simulation outcomes align with the theoretical findings presented in the previous sections.
The parameters considered here include
 $\psi=1$ and 2, $\varphi=0$, and $\sigma=1$ with sample sizes  $n=6$ and 10. For $n=6$, the values of $m$ are set to 1 and 2, and for $n=10$,  $m$ takes on the values of $1, 2, 3$ and 4. We perform $N=1000$ replications for the simulation. We derive the BLUEs and BLIEs of the location and scale parameters and  the BLUPs and BLIPs of $X_{q:n}$ where $q=n-m+1$. Assume $\widehat{Y}$ symbolizes a predictor of $Y$ and $\widehat{Y}_i$ is a prediction of $Y_i$, where $\widehat{Y}_i$ is derived in the $i$-th replication and $Y_i$ is the  real value of $Y$ generated at the same replication. Now, the estimated bias  (bias for short) and the estimated MSPE (EMSPE) of $\widehat{Y}$ are defined as follows
\begin{eqnarray*}
{\rm bias}(\widehat{Y})=\frac{1}{N}\sum_{i=1}^N(\widehat{Y}_i-Y_i),\quad\quad \quad {\rm and }\quad\quad\quad
{\rm EMSPE}(\widehat{Y})=\frac{1}{N}\sum_{i=1}^N(\widehat{Y}_i-Y_i)^2,
\end{eqnarray*}
respectively.

We compute and present the averages of the BLUEs and BLIEs for the location and scale parameters, along with the BLUPs and BLIPs for $X_{q:n}$,  and their biases and  EMSPEs  in Table 12.
Additionally, we compare the performance of the $95\%$ CIs and PIs using the average width (AW) and coverage probability (CP).
The results for $95\%$ CIs of the location and scale parameters and    $95\%$ PIs are reported in Table 13. Table 12 reveals that  1) the means of the  BLUEs are more closely aligned with the true parameter value than those of BLIEs, which is consistent with our expectations 3) The EMSPEs of the BLIEs are lower than those of the corresponding  BLUPs, which supports the previous theoretical findings.

\vspace{0.2cm}
From Table 13, it is observed that  1) the $95\%$ CIs for $\varphi$ based on $T_1$ and $T_3$ display identical performance, with one exception, 2)
the $95\%$ CIs for $\sigma$ based on $T_2$ and $T_4$ show very little difference in their performance,
3) the $95\%$ PIs based on $T^*_2$ exhibit smaller AWs compared to those based on $T_1^*$, whereas the corresponding CPs are rather close to each other.

\begin{table} [!htp]
\setlength{\tabcolsep}{0.44em}
\footnotesize
{\bf Table 12:} Averages of the BLUEs and BLIEs of $\varphi$ and $\sigma$ and BLUPs and BLIPs of $X_{q:n}$ ($Y$ for short) and  the biases and  EMSPEs for $XL(0,1,\psi)$.\vspace{-0.2cm}
	\begin{longtable}{|c|c|c|c|c|c|c|c|c|c|c|c|c|c|}
		\hline &&&&&&&&&&bias&bias &EMSPE&EMSPE\\
		$\psi$   &   $n$     & $m$ &$q$&$\widehat{\varphi}$ & $\widetilde\varphi$   &    $\widehat{\sigma}$ &     $\widetilde\sigma~$       &     $\widehat{Y}$ &     $\widetilde{Y}$  & $\widehat{Y}$ & $\widetilde{Y}$ & $\widehat{Y}$ & $\widetilde{Y}$ \\
\hline
1&	6&	1&	6&	0.0132&	0.0474&	0.7920&	0.6419&	2.35&	2.1670&	0.2991&	0.1161&	0.7240&	0.5655\\
&	&	2&	5&	0.0072&	0.0516&	0.8191&	0.6227&	1.4839&	1.3608&	0.0656&	-0.0575&0.2695&	0.2321\\
&	&	&	6&	0.0072&	0.0516&	0.8191&	0.6227&	2.4469&	2.0876&	0.3960&	0.0367&	1.4803&	1.0031\\
&	10&	1&	10&	0.0144&	0.0266&	0.8442&	0.7564&	2.934&	2.8276&	0.4212&	0.3148&	0.5331&	0.4332\\
&	&	2&	9&	0.0119&	0.0260&	0.8626&	0.7610&	2.0412&	1.9780&	0.0875&	0.0242&	0.2184&	0.2037\\
&	&	&	10&	0.0119&	0.0260&	0.8626&	0.7610&	3.0419&	2.8569&	0.5291&	0.3441&	1.0213&	0.7936\\
&	&	3&	8&	0.0094&	0.0259&	0.8809&	0.7614&	1.5793&	1.5289&	0.0519&	0.0015&	0.1178&	0.1091\\
&	&	&	9&	0.0094&	0.0259&	0.8809&	0.7614&	2.1037&	1.9797&	0.1499&	0.0260&	0.4481&	0.3921\\
&	&	&	10&	0.0094&	0.0259&	0.8809&	0.7614&	3.1251&	2.8591&	0.6123&	0.3464&	1.5230&	1.1181\\
&	&	4&	7&	0.0080&	0.0275&	0.8919&	0.7497&	1.2435&	1.1980&	0.0203&	-0.0251&0.0766&	0.0722\\
&	&	&	8&	0.0080&	0.0275&	0.8919&	0.7497&	1.6039&	1.4992&	0.0765&	-0.0283&0.2490&	0.2181\\
&	&	&	9&	0.0080&	0.0275&	0.8919&	0.7497&	2.1347&	1.9431&	0.1810&	-0.0106&0.6608&	0.5453\\
&	&	&	10&	0.0080&	0.0275&	0.8919&	0.7497&	3.1688&	2.8091&	0.6560&	0.2963&	2.0108&	1.3850\\
2&	6&	1&	6&	0.0105&	0.0246&	0.7526&	0.6039&	1.0091&	0.9268&	0.1432&	0.0609&	0.1489&	0.1167\\
&	&	2&	5&	0.0066&	0.0251&	0.7950&	0.5980&	0.6437&	0.5888&	0.0467&	-0.0082&0.0437&	0.0350\\
&	&	&	6&	0.0066&	0.0251&	0.7950&	0.5980&	1.0789&	0.9154&	0.2129&	0.0495&	0.2870&	0.1844\\
&	10&	1&	10&	0.0053&	0.0104&	0.8272&	0.7367&	1.3044&	1.2543&	0.1715&	0.1214&	0.1306&	0.1107\\
&	&	2&	9&	0.0042&	0.0101&	0.8472&	0.7426&	0.8969&	0.8678&	0.0437&	0.0146&	0.0444&	0.0409\\
&	&	&	10&	0.0042&	0.0101&	0.8472&	0.7426&	1.3588&	1.2721&	0.2259&	0.1392&	0.2296&	0.1818\\
&	&	3&	8&	0.0034&	0.0103&	0.8609&	0.7392&	0.6816&	0.6590&	0.0176&	-0.005&	0.0237&	0.0223\\
&	&	&	9&	0.0034&	0.0103&	0.8609&	0.7392&	0.9183&	0.8619&	0.0651&	0.0087&	0.0831&	0.0722\\
&	&	&	10&	0.0034&	0.0103&	0.8609&	0.7392&	1.3876&	1.2643&	0.2547&	0.1315&	0.3107&	0.2314\\
&	&	4&	7&	0.0024&	0.0106&	0.8786&	0.7334&	0.5378&	0.5175&	0.0147&	-0.0056&0.0145&	0.0134\\
&	&	&	8&	0.0024&	0.0106&	0.8786&	0.7334&	0.6996&	0.6523&	0.0356&	-0.0117&0.0468&	0.0410\\
&	&	&	9&	0.0024&	0.0106&	0.8786&	0.7334&	0.9410&	0.8536&	0.0878&	0.0004& 0.1293& 0.1049\\
&	&	&	10&	0.0024&	0.0106&	0.8786&	0.7334&	1.4199&	1.2529&	0.2871&	0.1201&	0.4200&	0.2916\\
3&	6&	1&	6&	0.0050&	0.0140&	0.7686&	0.6155&	0.6604&	0.6063&	0.0948&	0.0406&	0.0598&	0.0461\\
&	&	2&	5&	0.0031&	0.0149&	0.8006&	0.6010&	0.4122&	0.3768&	0.0226&	-0.0128&0.0198&	0.0169\\
&	&	&	6&	0.0031&	0.0149&	0.8006&	0.6010&	0.6943&	0.5884&	0.1287&	0.0228&	0.1141&	0.0750\\
&	10&	1&	10&	0.0023&	0.0055&	0.8255&	0.7342&	0.8245&	0.7922&	0.1147&	0.0824&	0.0510&	0.0424\\
&	&	2&	9&	0.0013&	0.0051&	0.8529&	0.7467&	0.5726&	0.5538&	0.0386&	0.0198&	0.0172&	0.0152\\
&	&	&	10&	0.0013&	0.0051&	0.8529&	0.7467&	0.8727&	0.8164&	0.1628&	0.1065&	0.0975&	0.0753\\
&	&	3&	8&	0.0007&	0.0051&	0.8698&	0.7459&	0.4360&	0.4213&	0.0139&	-0.0008&0.0096&0.0090\\
&	&	&	9&	0.0007&	0.0051&	0.8698&	0.7459&	0.5895&	0.5529&	0.0555&	0.0189&	0.0339&	0.0284\\
&	&	&	10&	0.0007&	0.0051&	0.8698&	0.7459&	0.8955&	0.8152&	0.1856&	0.1053&	0.1387&	0.1014\\
&	&	4&	7&	0.0001&	0.0054&	0.8856&	0.7384&	0.3419&	0.3288&	0.0084&	-0.0047&0.0064&	0.0059\\
&	&	&	8&	0.0001&	0.0054&	0.8856&	0.7384&	0.4462&	0.4158&	0.0241&	-0.0063&0.0200&	0.0174\\
&	&	&	9&	0.0001&	0.0054&	0.8856&	0.7384&	0.6025&	0.5460&	0.0685&	0.0120&	0.0536&	0.0418\\
&	&	&	10&	0.0001&	0.0054&	0.8856&	0.7384&	0.9141&	0.8057&	0.2042&	0.0958&	0.1885&	0.1288\\
4&	6&	1&	6&	0.0035&	0.0102&	0.7701&	0.6163&	0.4871&	0.4471&	0.0672&	0.0272&	0.0304&	0.0231\\
&	&	2&	5&	0.0024&	0.0110&	0.7947&	0.5962&	0.3000&	0.2742&	0.0128&	-0.013&	0.0117&	0.0100\\
&	&	&	6&	0.0024&	0.0110&	0.7947&	0.5962&	0.5062&	0.4289&	0.0863&	0.0089&	0.0635&	0.0418\\
&	10&	1&	10&	0.0010&	0.0034&	0.8137&	0.7234&	0.6006&	0.5772&	0.0867&	0.0632&	0.0267&	0.0221\\
&	&	2&	9&	0.0003&	0.0031&	0.8399&	0.7351&	0.4169&	0.4033&	0.0273&	0.0136&	0.0095&	0.0086\\
&	&	&	10&	0.0003&	0.0031&	0.8399&	0.7351&	0.6347&	0.5938&	0.1207&	0.0799&	0.0493&	0.0378\\
&	&	3&	8&	0.0000&	0.0032&	0.8529&	0.7312&	0.3158&	0.3052&	0.0079&	-0.0027&0.0050&	0.0048\\
&	&	&	9&	0.0000&	0.0032&	0.8529&	0.7312&	0.4265&	0.4001&	0.0368&	0.0104&	0.0175&	0.0149\\
&	&	&	10&	0.0000&	0.0032&	0.8529&	0.7312&	0.6476&	0.5897&	0.1336&	0.0757&	0.0682&	0.0493\\
&	&	4&	7&	0.0000&	0.0037&	0.8537&	0.7115&	0.2422&	0.2329&	0.0003&	-0.009&	0.0033&	0.0032\\
&	&	&	8&	0.0000&	0.0037&	0.8537&	0.7115&	0.3161&	0.2945&	0.0082&	-0.0133&0.0105&	0.0096\\
&	&	&	9&	0.0000&	0.0037&	0.8537&	0.7115&	0.4269&	0.3869&	0.0373&	-0.0028&0.0269&	0.0220\\
&	&	&	10&	0.0000&	0.0037&	0.8537&	0.7115&	0.6483&	0.5713&	0.1343&	0.0574&	0.0882&	0.0607\\
\hline
\end{longtable}
\end{table}

\begin{table} [!htp]
	\setlength{\tabcolsep}{0.36em}
	\footnotesize
	{\bf Table 13:} Average width and Coverage probabilities, for $XL(0,1,\psi)$.\vspace{-0.2cm}
\begin{longtable}{|c|c|c|c|cccccc|cccccc|}
\hline
$\psi$ & $n$ & $m$ & $q$ & \multicolumn{6}{c|}{Average Width} & \multicolumn{6}{c|}{Coverage Probability} \\ \cline{5-10}\cline{11-16}
& & & & $T_1$ & $T_2$ & $T_3$ & $T_4$ & $T_1^\ast$ & $T_2^\ast$ & $T_1$ & $T_2$ & $T_3$ & $T_4$ & $T_1^\ast$ & $T_2^\ast$ \\ \hline
1&	6&	1&	6&	0.9548&	2.3738&	0.9548&	2.3737&	5.5748&	4.5185&	0.930&	0.936&	0.930&	0.936&	0.943&	0.942\\
&	&	2&	5&	1.2614&	3.5134&	1.2614&	3.5132&	3.4676&	2.6361&	0.945&	0.951&	0.945&	0.951&	0.946&	0.942\\
&	&	&	6&	1.2614&	3.5134&	1.2614&	3.5132&	8.9030&	6.7682&	0.945&	0.951&	0.945&	0.951&	0.939&	0.939\\
&	10&	1&	10&	0.5340&	1.4652&	0.5341&	1.4652&	4.3934&	3.9367&	0.942&	0.936&	0.942&	0.936&	0.962&	0.965\\
&	&	2&	9&	0.5432&	1.6441&	0.5432&	1.6441&	2.4874&	2.1944&	0.940&	0.948&	0.940&	0.948&	0.958&	0.954\\
&	&	&	10&	0.5432&	1.6441&	0.5432&	1.6441&	5.6851&	5.0155&	0.940&	0.948&	0.940&	0.948&	0.946&	0.952\\
&	&	3&	8&	0.5799&	1.8642&	0.5799&	1.8643&	1.7409&	1.5048&	0.944&	0.946&	0.944&	0.946&	0.942&	0.938\\
&	&	&	9&	0.5799&	1.8642&	0.5799&	1.8643&	3.5865&	3.1001&	0.944&	0.946&	0.944&	0.946&	0.955&	0.951\\
&	&	&	10&	0.5799&	1.8642&	0.5799&	1.8643&	6.9970&	6.0482&	0.944&	0.946&	0.944&	0.946&	0.943&	0.953\\
&	&	4&	7&	0.6459&	2.2615&	0.6459&	2.2617&	1.5206&	1.2781&	0.948&	0.957&	0.948&	0.957&	0.954&	0.941\\
&	&	&	8&	0.6459&	2.2615&	0.6459&	2.2617&	2.6303&	2.2109&	0.948&	0.957&	0.948&	0.957&	0.949&	0.941\\
&	&	&	9&	0.6459&	2.2615&	0.6459&	2.2617&	4.6472&	3.9062&	0.948&	0.957&	0.948&	0.957&	0.950&	0.948\\
&	&	&	10&	0.6459&	2.2615&	0.6459&	2.2617&	8.1863&	6.8810&	0.948&	0.957&	0.948&	0.957&	0.947&	0.950\\
2&	6&	1&	6&	0.4234&	2.4838&	0.4234&	2.4838&	2.4241&	1.9452&	0.936&	0.934&	0.936&	0.934&	0.955&	0.959\\
&	&	2&	5&	0.5481&	3.4951&	0.5481&	3.4950&	1.6197&	1.2183&	0.943&	0.944&	0.943&	0.944&	0.963&	0.958\\
&	&	&	6&	0.5481&	3.4951&	0.5481&	3.4950&	4.0698&	3.0612&	0.943&	0.944&	0.943&	0.944&	0.947&	0.956\\
&	10&	1&	10&	0.2178&	1.4407&	0.2178&	1.4407&	2.0821&	1.8542&	0.943&	0.933&	0.943&	0.933&	0.948&	0.946\\
&	&	2&	9&	0.2233&	1.6061&	0.2233&	1.6061&	1.1211&	0.9827&	0.935&	0.930&	0.935&	0.929&	0.945&	0.947\\
&	&	&	10&	0.2233&	1.6061&	0.2233&	1.6061&	2.7398&	2.4017&	0.935&	0.930&	0.935&	0.929&	0.937&	0.941\\
&	&	3&	8&	0.2446&	1.8860&	0.2446&	1.8859&	0.7709&	0.6620&	0.947&	0.953&	0.947&	0.953&	0.934&	0.925\\
&	&	&	9&	0.2446&	1.8860&	0.2446&	1.8859&	1.6350&	1.4039&	0.947&	0.953&	0.947&	0.953&	0.966&	0.969\\
&	&	&	10&	0.2446&	1.8860&	0.2446&	1.8859&	3.1688&	2.7209&	0.947&	0.953&	0.947&	0.953&	0.943&	0.955\\
&	&	4&	7&	0.2666&	2.2018&	0.2666&	2.2018&	0.6655&	0.5556&	0.953&	0.948&	0.953&	0.948&	0.952&	0.951\\
&	&	&	8&	0.2666&	2.2018&	0.2666&	2.2018&	1.2485&	1.0422&	0.953&	0.948&	0.953&	0.948&	0.955&	0.955\\
&	&	&	9&	0.2666&	2.2018&	0.2666&	2.2018&	2.1234&	1.7726&	0.953&	0.948&	0.953&	0.948&	0.945&	0.956\\
&	&	&	10&	0.2666&	2.2018&	0.2666&	2.2018&	3.9378&	3.2871&	0.953&	0.948&	0.953&	0.948&	0.945&	0.952\\
3&	6&	1&	6&	0.2798&	2.3992&	0.2798&	2.3995&	1.6665&	1.3346&	0.941&	0.937&	0.941&	0.937&	0.941&	0.945\\
&	&	2&	5&	0.3496&	3.6315&	0.3496&	3.6317&	0.9740&	0.7312&	0.951&	0.963&	0.951&	0.963&	0.954&	0.947\\
&	&	&	6&	0.3496&	3.6315&	0.3496&	3.6317&	2.6200&	1.9668&	0.951&	0.963&	0.951&	0.963&	0.949&	0.952\\
&	10&	1&	10&	0.1339&	1.4540&	0.1339&	1.4540&	1.3167&	1.1711&	0.959&	0.932&	0.959&	0.932&	0.962&	0.962\\
&	&	2&	9&	0.1434&	1.6110&	0.1434&	1.6109&	0.7595&	0.6650&	0.952&	0.931&	0.952&	0.931&	0.954&	0.957\\
&	&	&	10&	0.1434&	1.6110&	0.1434&	1.6109&	1.7906&	1.5677&	0.952&	0.931&	0.952&	0.931&	0.953&	0.956\\
&	&	3&	8&	0.1589&	1.9312&	0.1589&	1.9314&	0.5197&	0.4457&	0.961&	0.953&	0.961&	0.953&	0.953&	0.949\\
&	&	&	9&	0.1589&	1.9312&	0.1589&	1.9314&	1.0608&	0.9098&	0.961&	0.953&	0.961&	0.953&	0.949&	0.954\\
&	&	&	10&	0.1589&	1.9312&	0.1589&	1.9314&	2.0789&	1.7829&	0.961&	0.953&	0.961&	0.953&	0.944&	0.955\\
&	&	4&	7&	0.1752&	2.2249&	0.1752&	2.2250&	0.4169&	0.3476&	0.958&	0.946&	0.958&	0.946&	0.946&	0.937\\
&	&	&	8&	0.1752&	2.2249&	0.1752&	2.2250&	0.8036&	0.6700&	0.958&	0.946&	0.958&	0.946&	0.953&	0.949\\
&	&	&	9&	0.1752&	2.2249&	0.1752&	2.2250&	1.3313&	1.1100&	0.958&	0.946&	0.958&	0.946&	0.939&	0.935\\
&	&	&	10&	0.1752&	2.2249&	0.1752&	2.2250&	2.5249&	2.1052&	0.958&	0.946&	0.958&	0.946&	0.927&	0.947\\
4&	6&	1&	6&	0.1960&	2.4017&	0.1960&	2.4014&	1.1909&	0.9531&	0.937&	0.930&	0.937&	0.930&	0.952&	0.951\\
&	&	2&	5&	0.2588&	3.5588&	0.2588&	3.5587&	0.7468&	0.5603&	0.946&	0.944&	0.946&	0.944&	0.951&	0.937\\
&	&	&	6&	0.2588&	3.5588&	0.2588&	3.5587&	1.9510&	1.4638&	0.946&	0.944&	0.946&	0.944&	0.954&	0.957\\
&	10&	1&	10&	0.0974&	1.4715&	0.0974&	1.4714&	0.9702&	0.8626&	0.934&	0.920&	0.934&	0.920&	0.950&	0.950\\
&	&	2&	9&	0.1050&	1.6413&	0.1050&	1.6413&	0.5212&	0.4561&	0.933&	0.921&	0.933&	0.921&	0.960&	0.957\\
&	&	&	10&	0.1050&	1.6413&	0.1050&	1.6413&	1.2709&	1.1123&	0.933&	0.921&	0.933&	0.921&	0.946&	0.956\\
&	&	3&	8&	0.1107&	1.8516&	0.1107&	1.8516&	0.3653&	0.3132&	0.940&	0.929&	0.940&	0.929&	0.963&	0.958\\
&	&	&	9&	0.1107&	1.8516&	0.1107&	1.8516&	0.7327&	0.6281&	0.940&	0.929&	0.940&	0.929&	0.949&	0.950\\
&	&	&	10&	0.1107&	1.8516&	0.1107&	1.8516&	1.5240&	1.3066&	0.940&	0.929&	0.940&	0.929&	0.951&	0.956\\
&	&	4&	7&	0.1218&	2.2089&	0.1218&	2.2092&	0.3025&	0.2522&	0.935&	0.936&	0.935&	0.937&	0.955&	0.948\\
&	&	&	8&	0.1218&	2.2089&	0.1218&	2.2092&	0.5754&	0.4796&	0.935&	0.936&	0.935&	0.937&	0.955&	0.949\\
&	&	&	9&	0.1218&	2.2089&	0.1218&	2.2092&	0.9307&	0.7757&	0.935&	0.936&	0.935&	0.937&	0.951&	0.956\\
&	&	&	10&	0.1218&	2.2089&	0.1218&	2.2092&	1.7496&	1.4583&	0.935&	0.936&	0.935&	0.937&	0.961&	0.970\\
\hline
\end{longtable}
\end{table}

\section{A Real Data Example}
The real data considered in this section are related to the fatigue life (in hours) for $10$ bearings of a certain type, due to \cite{McCool}. The data, arranged in order,  are:
\[
152.7,~172.0,~172.5,~173.3,~193.0,~204.7,~216.5,~234.9,~262.6,~422.6.
\]
We computed the coefficient of correlation between the data presented above and the means shown in Table 1 for $n=10$ and $\psi=4$. The result exceeded  $0.963$, allowing us to conclude that the data likely originates from the XLindley distribution with $\psi=4$.

\vspace{0.2cm}
We calculated the BLUEs and BLIEs of the location and scale parameters along with their respective MSEs expressed in terms of $\sigma^2$. Besides, we determined the predictions for $X_{10:10}$ and the MSPEs in terms of $\sigma^2$, see Table 14.
\begin{table}
\setlength{\tabcolsep}{0.5em}
\small
{\bf Table 14:} The BLUEs and BLIEs of the location and scale parameters and  the computed MSEs in terms of $\sigma^2$,
and the predictions of $X_{10:10}$ ($Y$ for short) and the MSPEs in terms of $\sigma^2$ for the example.\vspace{-0.2cm}
\begin{longtable}{|c|c|c|c|c|c|c|c|}
\hline
    &   $\widehat{\varphi}$ & $\widehat{\sigma}$   &  $\widetilde\varphi$   &     $\widetilde\sigma~$   &  MSE$(\widehat\varphi)$ & MSE$(\widehat\sigma)$ &$Cov(\widehat\varphi,\widehat\sigma)$\\
\hline
Complete Sample&145.1533  &289.8381& 145.9070 &   260.9310&0.0008&0.1108&-0.0029\\
$c=1$, $q=10$&146.2211 &248.8363 & 146.9404 &221.2461 & 0.0008&0.1247   &-0.0033  \\
\hline \hline &&&&&&&\\
 & MSE$(\widetilde\varphi)$	& MSE$(\widetilde\sigma)$	& $\widehat Y$ &     $\widetilde Y$	& $V_4$ 	& MSPE$(\widehat Y)$	&MSPE$(\widetilde Y)$\\
\hline
Complete Sample&0.0007&0.0997 &  &   &  &   &  \\
$c=1$, $q=10$& 0.0008&0.1109 &327.1256  &  319.9548 &  0.0324 &  0.0755  &0.0745  \\
\hline
\end{longtable}
\end{table}
\newpage
\section{Conclusion}
We studied the order statistics from the XLindley distribution  and developed  formulas for both the single and product moments of these order statistics. We proceeded to calculate the means, variances, and covariances of order statistics for specific cases.  Following that, our focus shifted to deriving the BLUEs and BLIEs of the location and scale parameters. We  focused on the methodologies for obtaining the BLUP and BLIP of a future unobserved order statistic. 
We additionally explored the construction of confidence intervals for both the location and scale parameters, alongside prediction intervals for an unobserved order statistic. Additionally, we conducted a simulation study and included an example. The  results highlight that the BLIE  outperforms the BLUE in terms of MSE and
the BLIP  surpasses the  BLUP  in terms of MSPE. Furthermore, prediction intervals based on the BLIE of the scale parameter showed better performance compared to those derived based on the BLUE of the scale parameter in terms of AW. In terms of confidence intervals, those for the location parameter exhibited approximately identical performance, while the confidence intervals for the scale parameter showed close performance in relation to both AW and CP.
 All computations were executed utilizing the R software   \cite{RC}.

\vspace{0.25cm}
The findings of this paper rely on the premise that the shape parameter $\psi$ is known. This presumption has been adopted by several other researchers, as well. 
Now, if this parameter is not known, we may substitute it with an appropriate estimate, allowing us to achieve the approximate BLUEs and BLIEs, as well as approximate BLUPs and BLIPs for future order statistics.\\

{\bf{Compliance with ethical standards}}\\

{\bf{Conflict of interest:}} On behalf of all authors, the corresponding author states that there is no conflict of
interest.\\


\begin{thebibliography}{99}
\bibitem{Ahsan}Ahsanullah, M. and Alzaatreh, A. (2018). Parameter estimation for the log-logistic distribution based on order statistics. {\em REVSTAT-Statistical Journal}, {\bf 16}, 429--443.
\bibitem{ZA1}Akhter Z., MirMostafaee, S.M.T.K. and Athar, H. (2019). On the moments of order statistics from the standard two-sided power distribution. {\em Journal of Mathematical Modeling},  7, 381-398.
\bibitem{ZA3} Akhter, Z., MirMostafaee, S.M.T.K. and Ormoz, E. (2022). On the order statistics of exponentiated moment exponential distribution and associated inference. {\em Journal of Statistical Computation and Simulation}, 92(6), 1322-1346.
\bibitem{ZA2}Akhter, Z., Saran, J., Verma, K. and Pushkarna, N. (2022).  Moments of order statistics from length-biased exponential distribution and associated inference. {\em Annals of Data Science},  9, 1257-1282.
\bibitem{Alo} Alotaibi, R., Nassar, M., and Elshahhat, A. (2022). Computational analysis of XLindley parameters using adaptive Type-II progressive hybrid censoring with applications in chemical engineering. {\em Mathematics}, 10(18), 3355.
\bibitem{arnd} Arnold, B.C., Balakrishnan, N., and Nagaraja, H.N. (2008). A First Course in Order Statistics. Society for Industrial and Applied Mathematics.
\bibitem{Bala}
Balakrishnan, N., and Chan, P.S. (1992). Order statistics from extreme value distribution, II: best linear unbiased estimates and some other uses. {\em Communications in Statistics-Simulation and Computation}, 21(4), 1219-1246.
\bibitem{BR} Balakrishnan, N., and Rao, C.R. (1998).  Order statistics: An introduction. Handbook of statistics, 16, 3-24.
\bibitem{Burk} Burkshat, M. (2010). Linear estimators and predictors based on generalized order statistics from generalized Pareto distributions,
{\em Communications in Statistics-Theory and Methods}, {\bf 39},  311--326.
\bibitem{Cen}\c{C}etinkayam, \c{C}. and Gen\c{c}, A.\.{l}. (2018). Moments of order statistics of the standard two-sided power distribution. {\em Communications in Statistics - Theory and Methods},  47, 4311-4328.
\bibitem{Chouia}Chouia, S., and Zeghdoudi, H. (2021). The XLindley distribution: Properties and application. 
{\em Journal of Statistical Theory and Applications}, 20(2), 318-327.
\bibitem{david} David, H.A., and Nagaraja, H.N. (2003). Order Statistics. John Wiley and Sons.
\bibitem{Gen}Gen\c{c}, A.\.{l}. (2012). Moments of order statistics of Topp-Leone distribution. {\em Statistical Papers,}  53, 117--131.
\bibitem{GR}Gradshteyn, I.S. and Ryzhik, I.M. (2007). {\em Tables of Integrals, Series and Products}. Edited by Jeffrey, A. and Zwillinger, D., 7th Edn., Academic Press, San Diego.
\bibitem{Guan} Guan, R., Cheng, W., Rong, Y. and Zhao, X. (2023). Parameter estimation of Beta-Exponential distribution using linear combination of order statistics. {\em Commun. Math. Stat.}, https://doi.org/10.1007/s40304-022-00306-6    
\bibitem{hos} Hosking, J.R. (1990). L-moments: analysis and estimation of distributions using linear combinations of order statistics. {\em Journal of the Royal Statistical Society Series B: Statistical Methodology}, 52(1), 105-124.
\bibitem{KN} Kaminsky K.S., and Nelson P.I. (1975). Best linear unbiased prediction of order statistics in location and scale families.
{\em Journal of the Amirecan Statistical Association}, {70}(349), 145-150.
\bibitem{lind} Lindley, D.V. (1958). Fiducial distributions and Bayes' theorem. {\em Journal of the Royal Statistical Society. Series B (Methodological)}, 102-107.
\bibitem{Lloyd} Llyod E.H. (1952). Least squares estimation of location and scale parameters using order statistics. {\em Biometrika}, 39(1-2), 88-95.
\bibitem{Mah}Mahmoud, M.A.W., Sultan, K.S. and Amer, S.M. (2003). Order statistics from inverse Weibull distribution and associated inference. 
{\em Computational Statistics \& Data Analysis}, 42(1-2), 149-163.
\bibitem{Man} Mann, N.R. (1969). Optimum estimators for linear functions of location and scale parameters. {\em The Annals of  Mathematical Statistics},
40, 2149-2155.
\bibitem{McCool}McCool, J.L. (1974). Inferential Techniques for Weibull Populations. {\em Aerospace Research Laboratories Report ARL TR 74-0180}, Wright-Patterson AFB, Ohio.
\bibitem{Metri} Metiri, F., Zeghdoudi, H., and Ezzebsa, A. (2022). On the characterisation of X-Lindley distribution by truncated moments. Properties and application. {\em Operations Research and Decisions}, 32(1), 97-109.
\bibitem{mir}MirMostafaee, S.M.T.K. (2014). On the moments of order statistics coming from the Topp-Leone distribution. {\em Statistics and Probability Letters}, 95, 85--91.
\bibitem{Nad}Nadarajah, S. (2008). Explicit expressions for moments of order statistics. {\em Statistics and Probability Letters}, {78}, 196--205.
\bibitem{Nag}Nagaraja, H.N. (2013). Moments of order statistics and L-moments for the symmetric triangular distribution. {\em Statistics and Probability Letters},  83, 2357-2363.
\bibitem{Nsr} Nassar, M., Alotaibi, R., and Elshahhat, A. (2023). Reliability estimation of XLindley constant-stress partially accelerated life tests using progressively censored samples. {\em Mathematics}, 11(6), 1331.
\bibitem{RC}R Core Team (2022). R: A Language and Environment for Statistical Computing. R Foundation for Statistical Computing, Vienna, Austria.
  http://www.R-project.org.
\bibitem{rah}Raqab, M.M. and Ahsanullah, M. (2001). Estimation of the location and scale parameters of generalized exponential distribution based on order statistics. {\em Journal of Statistical Computation and Simulation},  69, 109-123.
\bibitem{shah} Shahbaz, M. Q., Ahsanullah, M., Shahbaz, S.H. and Al-Zahrani, B.M. (2016). Ordered random variables: Theory and applications.
\bibitem{snker1}
Shanker, R. and Mishra, A. (2013). A quasi Lindley distribution. African Journal of Mathematics and Computer Science Research, 6(4), 64-71.
\bibitem{snker2}
Shanker, R., Sharma, S. and Shanker, R. (2013). A two-parameter Lindley distribution for modeling waiting and survival times data. {\em Applied Mathematics}, 4(2) 363–368.
\bibitem{sal}Sultan, K.S. and AL-Thubyani, W.S. (2016). Higher order moments of order statistics from the Lindley distribution and associated inference. {\em Journal of Statistical Computation and Simulation},  86, 3432-3445.
\bibitem{Zanjiran}
Zanjiran, F. and MirMostafaee, S.M.T.K. (2024). On estimation and prediction for the XLindley distribution based on record data.
{\em  Reliability: Theory \& Applications},  19(2), 258-272.

\end{thebibliography}
\end{document}